\documentclass[showpacs,aps,prd,preprint,nofootinbib,showkeys,unsortedaddress,raggedbottom]{revtex4-1}
\pdfoutput=1

\usepackage{graphicx}
\usepackage{amsmath}
\usepackage{amssymb}

\usepackage[usenames,dvipsnames]{color}
\definecolor{darkblue}{RGB}{0,0,196}
\usepackage[colorlinks=true,linkcolor=darkblue,citecolor=darkblue,urlcolor=darkblue]{hyperref}

\def\ba{\begin{eqnarray}}
\def\ea{\end{eqnarray}}

\def\be{\begin{equation}}
\def\ee{\end{equation}}

\newcommand{\checked}[1]{}

\begin{document}

\title{The anisotropic non-equilibrium hydrodynamic attractor}

\author{Michael Strickland} 
\affiliation{Department of Physics, Kent State University, Kent, OH 44242 United States}

\author{Jorge Noronha} 
\affiliation{Instituto de F\'isica, Universidade de S\~ao Paulo, Rua do Mat\~ao, 1371, Butant\~a, 05508-090, S\~ao Paulo, SP, Brazil}

\author{Gabriel S.~Denicol} 
\affiliation{Instituto de F\'isica, Universidade Federal Fluminense, UFF, Niter\'oi, 24210-346, RJ, Brazil}

\begin{abstract}
We determine the dynamical attractors associated with anisotropic hydrodynamics (aHydro) and the DNMR equations for a 0+1d conformal system using kinetic theory in the relaxation time approximation.  We compare our results to the non-equilibrium attractor obtained from exact solution of the 0+1d conformal Boltzmann equation, Navier-Stokes theory, and second-order Mueller-Israel-Stewart theory.  We demonstrate that the aHydro attractor equation resums an infinite number of terms in the inverse Reynolds number.  The resulting resummed aHydro attractor possesses a positive longitudinal to transverse pressure ratio and is virtually indistinguishable from the exact attractor. This suggests that kinetic theory involves not only a resummation in gradients (Knudsen number) but also a novel resummation in inverse Reynolds number.  We also demonstrate that the DNMR result provides a better approximation to the exact kinetic theory attractor than Mueller-Israel-Stewart theory.  Finally, we introduce a new method for obtaining approximate aHydro equations which relies solely on an expansion in inverse Reynolds number, carry out this expansion to third order, and compare these third-order results to the exact kinetic theory solution.
\end{abstract}

\date{\today}


\maketitle

\section{Introduction}
\label{sec:intro}

Relativistic hydrodynamics is currently the main theoretical approach to describe the time evolution of the rapidly expanding quark-gluon plasma (QGP) produced in ultrarelativistic heavy ion collisions \cite{Heinz:2013th}. However, despite its success, understanding how hydrodynamics can provide a reasonable description of the rapidly expanding matter formed in these collisions is not an easy task. Traditionally, hydrodynamics has been understood as a truncation of a gradient expansion \cite{Chapman_Enskog} and, thus, its domain of validity could only be justified when gradients were sufficiently smooth when compared to the inverse microscopic scales of the problem. In fact, the gradient expansion itself was previously understood as a universal macroscopic limit displayed by microscopic theories, reached at sufficiently late times. However, it has been recently shown \cite{Heller:2013fn,Buchel:2016cbj,Heller:2016rtz,Denicol:2016bjh} that the gradient expansion has zero radius of convergence for flow configurations that are relevant for the QGP (both at strong coupling and also in kinetic theory) and, in this sense, one cannot construct and improve a hydrodynamic theory by systematically taking into account higher order terms in this series. Therefore, the concept that relativistic hydrodynamics is only applicable when gradients of macroscopic quantities are small, derived from the gradient expansion, seems to be no longer well justified (or even needed). In the end, these findings have lead one to revisit the very definition of viscous hydrodynamics in order to assess its domain of applicability in heavy ion collisions.

As a matter of fact, though the early success of fluid-dynamical models was
initially interpreted as a signature of rapid thermalization of the quark-gluon
plasma \cite{Heinz:2004qz}, model calculations \cite{Chesler:2008hg,Beuf:2009cx,Chesler:2009cy,Heller:2011ju,Heller:2012je,Heller:2012km,vanderSchee:2012qj,Casalderrey-Solana:2013aba,vanderSchee:2013pia,Heller:2013oxa,Keegan:2015avk,Chesler:2015bba,Kurkela:2015qoa,Chesler:2016ceu,Attems:2016ugt,Attems:2016tby,Attems:2017zam,Florkowski:2017olj} have suggested that such interpretation was premature given that systems far from equilibrium may already display hydrodynamic behavior via a process known as \emph{hydrodynamization}, a novel feature of rapidly expanding fluids such as the QGP. Naturally, the validity of hydrodynamics is not without bounds: it will eventually fail when the values of viscosity become sufficiently large or when it is applied at sufficiently early times. Nevertheless, even in such extreme cases, it is possible to devise effective theories that are capable of describing the quark-gluon plasma, the most notable being anisotropic
hydrodynamics (aHydro) \cite{Florkowski:2010cf,Martinez:2010sc,Ryblewski:2010ch,Martinez:2012tu,Ryblewski:2012rr,Bazow:2013ifa,Tinti:2013vba,Nopoush:2014pfa,Tinti:2015xwa,Bazow:2015cha,Strickland:2015utc,Alqahtani:2015qja,Molnar:2016vvu,Molnar:2016gwq,Alqahtani:2016rth,Bluhm:2015raa,Bluhm:2015bzi,Alqahtani:2017jwl,Alqahtani:2017tnq}. 

In general, hydrodynamization is now expected to occur at a time scale $\tau_{\rm hydro}$ shorter than the corresponding time scales for isotropization and thermalization, driven by a novel \emph{dynamical attractor} whose details vary according to the theory under consideration, e.g., kinetic theory, hydrodynamics, holography and etc \cite{Heller:2015dha,Keegan:2015avk,Romatschke:2017vte,Bemfica:2017wps,Spalinski:2017mel}. Such attractor solutions show that hydrodynamics displays a new degree of universality far-from-equilibrium regardless of the details of the initial state of the system. In fact, the approach to the dynamical attractor effectively wipes out information about the specific initial condition used for the evolution, before the true equilibrium state and consequently, thermalization, is reached. 

In the context of kinetic theory and standard statistical mechanics, thermalization is understood as the development of isotropic thermal one-particle distribution functions for the partons which comprise the QGP.  In a high-energy heavy-ion collision, the large longitudinal expansion rate causes the center of the QGP fireball to only slowly relax to an approximately isotropic state with $\tau_{\rm iso} \gtrsim 3-4$ fm/c \cite{Strickland:2013uga}\footnote{We note also that studies of non-equilibrium QGP dynamics using either the 2PI formalism or holography indicate that, in the highest temperatures probed during heavy-ion collisions, an equation of state may be established well before pressure isotropization occurs \cite{Berges:2004ce,Attems:2016ugt,Attems:2017zam}.}; however, the time scale for hydrodynamization of the fireball appears to be much shorter (for a review see \cite{Florkowski:2017olj}). The catch, however, is that in practice one finds that the relevant quantity for judging whether one is close to attractor behavior is the dimensionless variable $w \equiv \tau T$ \cite{Heller:2015dha} which, in conformal fluids undergoing Bjorken expansion \cite{Bjorken:1982qr}, is proportional to the inverse of the Knudsen number $K_N$ with $1/T$ being the microscopic time scale. For small gradients where $w > 1$, the system follows the dynamics consistent with the dynamical attractor. However, in the large gradient regime where $w \ll 1$ the dynamics of the system is dominated by non-hydrodynamic modes (i.e., modes in the linearized dynamics whose frequency remains nonzero even for a spatially homogeneous system \cite{Kovtun:2005ev}) whose evolution depends on the precise initial condition assumed.  If we consider a fixed proper time after the collision, this implies that as we move closer to the edge of the QGP one will be more sensitive to the truly non-equilibrium dynamics associated with non-hydrodynamic modes. As a consequence, some non-universal aspects of the underlying theory, be they e.g. kinetic theory or holographically inspired, will start to affect the spatiotemporal evolution of the QGP. In this case, one must make a choice as to which underlying microscropic theory best reflects the relevant physics.  Since, as one moves close to the QGP edge, the system is much more dilute, a kinetic theory approach would seem to be preferred in this spatial region.

For this reason, in this paper we investigate the dynamical attractors of different approximations to the relativistic Boltzmann equation.  We determine the dynamical attractors associated with aHydro and Denicol-Niemi-Molnar-Rischke (DNMR) effective theory \cite{Denicol:2012cn} for 0+1d conformal kinetic theory in the relaxation time approximation \cite{anderson1974relativistic}. We compare our results for the non-equilibrium attractor in these theories with the corresponding results obtained from the exact solution of the 0+1d conformal Boltzmann equation and also second-order Mueller-Israel-Stewart (MIS) theory \cite{Muller:1967zza,Israel:1976tn,Israel:1979wp}.  In this paper, we show for the first time that the aHydro formalism has an attractor solution which, surprisingly, is in very good agreement with the attractor solution of corresponding microscopic theory. We further demonstrate that, in the aHydro formalism, the equation of motion for the shear stress tensor involves a resummation of an infinite number of terms in the inverse Reynolds number \cite{Denicol:2012cn}. Such terms are not present in traditional hydrodynamic theories and we consider that this novel feature is the  main reason behind the optimal agreement between the attractors of aHydro and those of the Boltzmann equation (in the relaxation time approximation).

This suggests that kinetic theory involves not only a resummation in gradients (Knudsen number) but also a novel resummation in inverse Reynolds number. Correspondingly, we also demonstrate that the DNMR result provides a better approximation to the exact kinetic theory attractor than MIS theory.  Finally, we introduce a new method for obtaining approximate aHydro equations which relies solely on an expansion in inverse Reynolds number, carry out this expansion to third order, and compare the third-order results to the exact solution.

This paper is structured as follows. In the next section we define the kinetic theory model used and the corresponding second order hydrodynamic theories we consider in this work. Anisotropic hydrodynamics is discussed in Section \ref{aHydrosection}. We investigate the attractor behavior of the different models in Section \ref{sec:attractorvars}, while numerical results can be found in Section \ref{results}. We finish with our conclusions and outlook in Section \ref{conclusions}. Appendices \ref{appendix1} and \ref{tinti} are included to further investigate different approximations and prescriptions within anisotropic hydrodynamics. 

\section{Kinetic theory and second order hydrodynamics}

We assume that the system is 0+1d, i.e.~transversally homogeneous and boost-invariant \cite{Bjorken:1982qr}.  As a result all variables will only depend on the longitudinal proper time, $ \tau = \sqrt{t^2-z^2}$.  The metric is taken to be ``mostly minus'' with $x^\mu = (t, x, y, z)$,  where the line element is $ds^2=g_{\mu\nu} dx^\mu dx^\nu=dt^2-dx^2-dy^2-dz^2$ with $g^{\mu\nu}$ being metric tensor in Minkowski space. The longitudinal spacetime rapidity is $\varsigma = \tanh^{-1} (z/t)$.  We assume that the system is conformal \cite{Baier:2007ix} with an equation of state corresponding to $N_{\rm dof}$ massless degrees of freedom which is Landau-matched \cite{LandauLifshitzFluids} to the general non-equilibrium energy density, i.e. $\epsilon_0(T) = \epsilon$.  In this case, one has $\epsilon = \epsilon_0(T) = 3 P_0(T)$ and $T = \gamma \epsilon^{1/4}$, where $\gamma$ is proportional to $N_{\rm dof}^{-1/4}$. Also, for a (Bjorken) longitudinally boost-invariant system the flow velocity is \mbox{$u^\mu = (\cosh\varsigma,0,0,\sinh\varsigma)$}.

We will use kinetic theory to obtain the aHydro and second-order viscous hydrodynamics dynamical attractors.  For this purpose we start from the Boltzmann equation in the relaxation time approximation (RTA) \cite{anderson1974relativistic}
\be
p^\mu\partial_\mu f = - \frac{p_\mu u^\mu}{\tau_{\rm eq}} \left( f - f_{\rm eq}\right) \, .
\label{BoltzmannRTA}
\ee
where $\tau_{\rm eq} = 5 \eta/(sT)$ \cite{Denicol:2010xn,Denicol:2011fa} is the relaxation time with $\eta$ being the shear viscosity, $T$ is the local effective temperature obtained via Landau matching, and $s$ is the entropy density. For this massless gas, the Boltzmann RTA equation changes covariantly under conformal transformations \cite{Denicol:2014xca,Denicol:2014tha} and $\eta/s$ is constant. We will assume classical Boltzmann distributions throughout, i.e. the equilibrium distribution function is $f_{\rm eq}(x) = \exp(-x)$.

In kinetic theory the covariantly conserved energy-momentum tensor is given by
\be
T^{\mu\nu} = N_{\textrm{dof}}\int dP \,p^{\mu} p^\nu \,f,
\ee
with $\int dP$ being the appropriate Lorentz invariant measure \cite{anderson1974relativistic}. The local energy density is obtained via $\epsilon = u_\mu u_\nu T^{\mu\nu}$ whereas the shear stress tensor is 
\be
\Pi^{\mu\nu} = \Delta^{\mu\nu}_{\alpha\beta}T^{\alpha\beta},
\ee
where $\Delta^{\mu\nu}_{\alpha\beta} = \left(\Delta^{\mu}_\alpha\Delta^{\nu}_\beta + \Delta^{\mu}_\beta\Delta^{\nu}_\alpha\right)/2 - \Delta^{\mu\nu}\Delta_{\alpha\beta}/3$ is the tensor projector orthogonal to the flow constructed using $\Delta_{\mu\nu} = g_{\mu\nu}-u_\mu u_\nu$.

Bjorken symmetry and conformal invariance may be used to show that the energy-momentum conservation laws, obtained from the first moment of the Boltzmann equation, can be reduced to a single equation 
\be
\tau \frac{d\log \epsilon}{d\tau}  = -\frac{4}{3} + \frac{\Pi}{\epsilon}
\label{eq:firstmom}
\ee 
involving the energy density and $\Pi = \Pi^\varsigma_\varsigma$.  In second-order hydrodynamic theories, such as MIS \cite{Muller:1967zza,Israel:1976tn,Israel:1979wp} and DNMR \cite{Denicol:2012cn,Denicol:2014loa}, one uses the 14-moment approximation 
for the single particle distribution function to obtain the most simple form of a differential equation for $\Pi$, which can be written in the following form
\be
\dot\Pi = \frac{4\eta}{3\tau\tau_\pi} - \beta_{\pi\pi}\frac{\Pi}{\tau} - \frac{\Pi}{\tau_\pi},
\label{2ndorderhydro}
\ee
where $\,\dot{}\,=d/d\tau$ and for RTA $\beta_{\pi\pi} = 38/21$ and $\tau_\pi = \tau_{\rm eq}$ in the complete second order calculation (which is the case for DNMR) \cite{Denicol:2010xn,Denicol:2012cn,Denicol:2014loa,Jaiswal:2013vta,Jaiswal:2013npa}, while in MIS $\beta_{\pi\pi} = 4/3$ and $\tau_\pi = 6\tau_{\rm eq}/5$ \cite{Muronga:2003ta}. By solving Eqs.\ \eqref{eq:firstmom} and \eqref{2ndorderhydro} one can determine the dynamical evolution of this viscous fluid described by second order hydrodynamics and investigate the emergence of hydrodynamic attractor behavior, as done in \cite{Heller:2015dha}. 

\section{Anisotropic hydrodynamics} 
\label{aHydrosection}   

The formalism behind anisotropic hydrodynamics has been explored in a series of papers (see e.g. \cite{Florkowski:2010cf,Martinez:2010sc,Bazow:2013ifa,Tinti:2013vba,Nopoush:2014pfa,Tinti:2015xwa,Bazow:2015cha,Alqahtani:2015qja,Molnar:2016vvu,Molnar:2016gwq}) and we refer the reader to those references for details. Here we only present the main points needed in this paper to make the discussion self-consistent. 

In the 0+1d case aHydro requires only one anisotropy direction and parameter, $\hat{\bf n}$ and $\xi$.  This leads to a distribution function Ansatz of the form \cite{Romatschke:2003ms,Romatschke:2004jh}
\be
f(\tau,{\bf x},{\bf p}) = f_{\rm eq} \!\left( \frac{1}{\Lambda(\tau,{\bf x})} \sqrt{ p_T^2 + [1+\xi(\tau,{\bf x})] p_L^2 } \right) ,
\ee
where $\Lambda$ can be interpreted as the local ``transverse temperature''.
For a conformal system, using this form, one finds that the energy density, transverse pressure, and longitudinal pressure factorize, resulting in
\ba
\epsilon &=& {\cal R}(\xi) \epsilon_0(\Lambda) \, , \nonumber \\
{\cal P}_T &=& {\cal R}_T(\xi) P_0(\Lambda) \, , \nonumber \\
{\cal P}_L &=& {\cal R}_L(\xi) P_0(\Lambda) \, , \nonumber
\ea
with \cite{Rebhan:2008uj,Martinez:2010sc}
\ba
{\cal R}(\xi) &=& \frac{1}{2}\left[\frac{1}{1+\xi}
+\frac{\arctan\sqrt{\xi}}{\sqrt{\xi}} \right] ,
\label{eq:rfunc}
\\
{\cal R}_{T}(\xi) &=& \frac{3}{2 \xi} 
\left[ \frac{1+(\xi^2-1){\cal R}(\xi)}{\xi + 1}\right] ,
\\
{\cal R}_{L}(\xi) &=& \frac{3}{\xi} 
\left[ \frac{(\xi+1){\cal R}(\xi)-1}{\xi+1}\right] ,
\ea
which satisfy $3{\cal R} = 2 {\cal R}_T + {\cal R}_L$ (the isotropic pressure is $P_0 = \epsilon/3$).  In all expressions above, $L$ and $T$ correspond to the directions parallel and perpendicular to $\hat{\bf n}$, respectively.  Conventionally, the anisotropy direction is taken to point in the beam line direction in heavy-ion applications ($\hat{\bf n} = \hat{\bf z}$).  Using Landau matching, one has $\epsilon = {\cal R}(\xi) \epsilon_0(\lambda) = \epsilon_0(T)$, which results in 
\be
T = {\cal R}^{1/4}(\xi) \Lambda.
\label{defineLambda}
\ee
Now we need an equation of motion for $\xi$ since $\Lambda$ is already connected to the temperature via the equation above. 

We also employ the following moment of the Boltzmann distribution \cite{Nopoush:2014pfa}
\be
I^{\mu\nu\lambda} = N_{\textrm{dof}}\int dP\, p^{\mu}p^\nu p^\lambda\,f,
\ee
which will be important for the aHydro approach. Using the Boltzmann equation in the RTA \eqref{BoltzmannRTA}, the equation of motion for this moment is
\be
\partial_\alpha I^{\alpha\mu\nu} = \frac{1}{\tau_{\rm eq}} ( u_\alpha I^{\alpha\mu\nu}_{\rm eq} - u_\alpha I^{\alpha\mu\nu} ) \, .
\label{eq:secondmom1}
\ee
We note that $I^{\mu\nu\lambda}$ is symmetric with respect to interchanges of $\mu$, $\nu$, and $\lambda$ and traceless in any pair of indices (massless particles/conformal invariance).  In an isotropic system, one finds $I_{xxx} = I _{yyy} = I_{zzz} = I_0$ with
\be
I_0(\Lambda) = \frac{4 N_{\rm dof}}{\pi^2} \Lambda^5 \, .
\ee
Using the aHydro form one finds
\ba
I_{uuu} &=& \frac{3+2\xi}{(1+\xi)^{3/2}} I_0(\Lambda) \, , \nonumber \\
I_{xxx} = I_{yyy} &=& \frac{1}{\sqrt{1+\xi}} I_0(\Lambda) \, , \nonumber \\
I_{zzz} &=& \frac{1}{(1+\xi)^{3/2}} I_0(\Lambda) \, ,
\ea
with, e.g. $I_{uuu} \equiv u_\mu u_\nu u_\lambda I^{\mu\nu\lambda}$, etc.

Taking the $zz$ projection of Eq.~(\ref{eq:secondmom1}) minus one-third of the sum of its $xx$, $yy$, and $zz$ projections gives our second equation of motion
\be
\frac{1}{1+\xi} \dot\xi - \frac{2}{\tau} + \frac{{\cal R}^{5/4}(\xi)}{\tau_{\rm eq}} \xi \sqrt{1+\xi} = 0\, ,
\label{eq:2ndmomf}
\ee
which can be used to define the evolution of the anisotropy parameter. 

\subsection{Connection with shear stress tensor and the inverse Reynolds number}

\begin{figure}[t!]
\centerline{
\includegraphics[width=.45\linewidth]{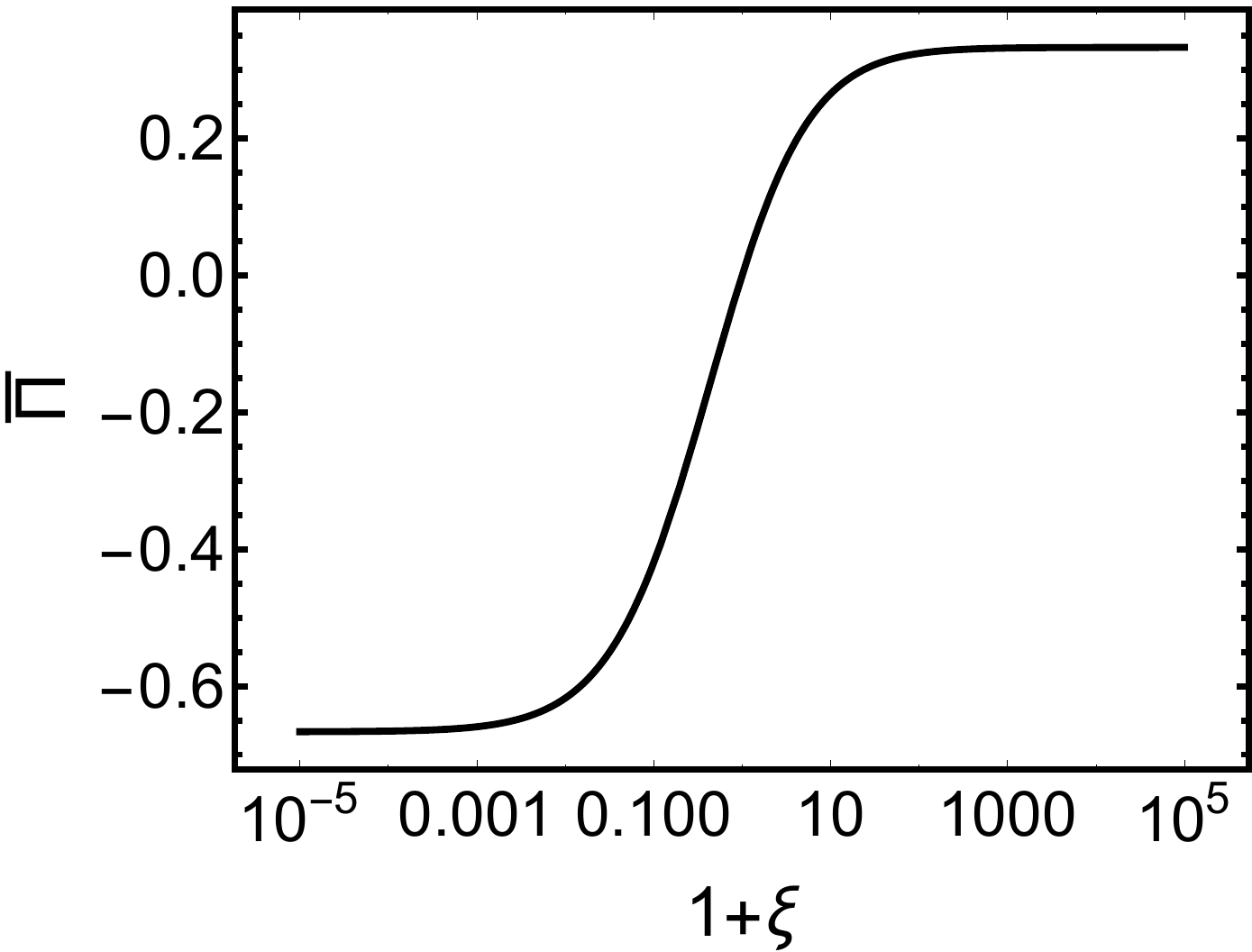}
\hspace{5mm}
\includegraphics[width=.45\linewidth]{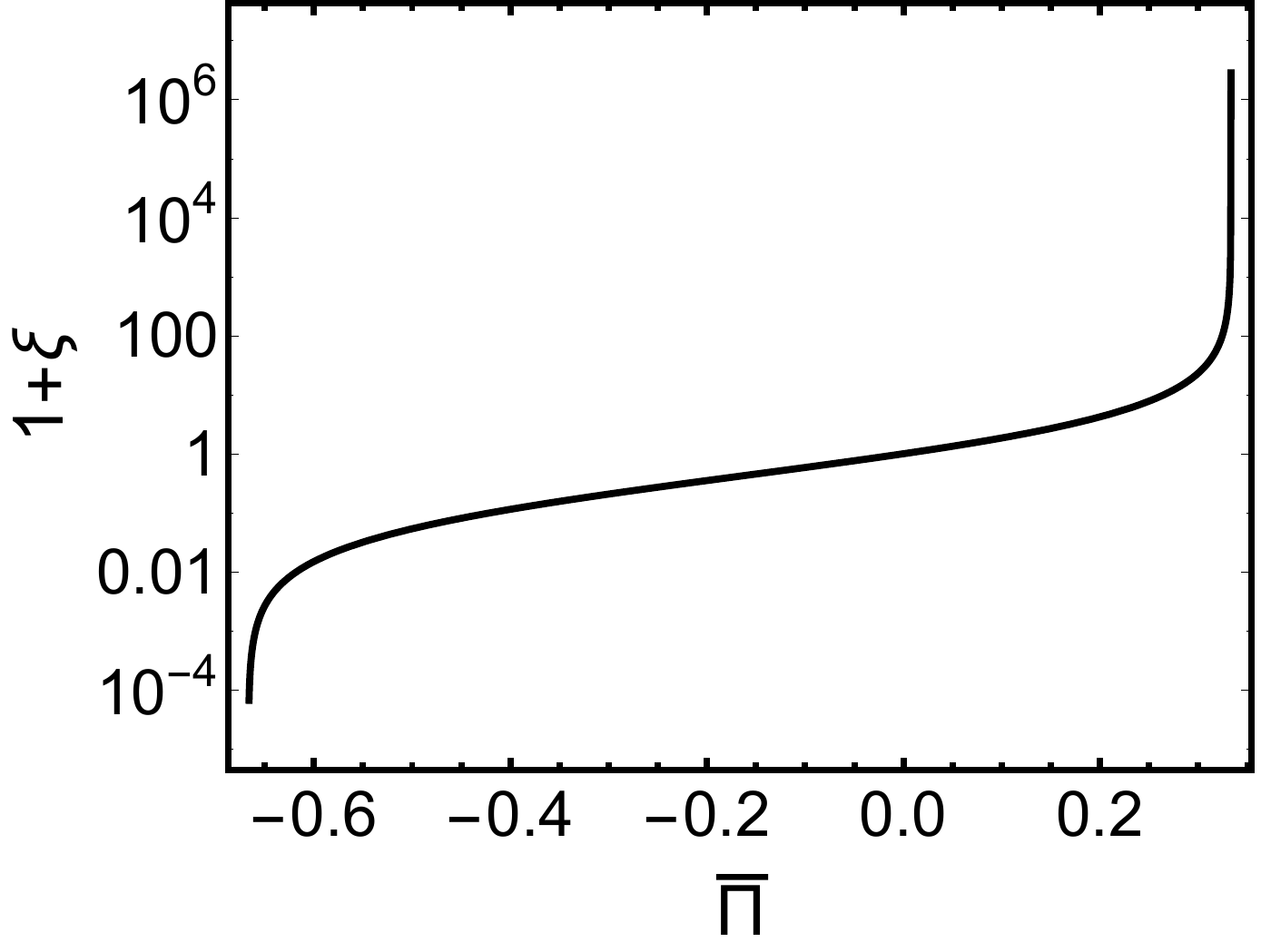}
}
\caption{The left panel shows $\overline{\Pi}$ as a function of $\xi$ determined via Eq.~(\ref{eq:pixirel}).  The right panel shows $\xi$ as a function of $\overline{\Pi}$ determined via numerical inversion of Eq.~(\ref{eq:pixirel}).}
\label{fig:pibar}
\end{figure}

In order to proceed in a manner that will allow a more transparent comparison between the aHydro equations of motion and those of standard viscous hydrodynamics, we will rewrite Eq.~(\ref{eq:2ndmomf}) in terms of the shear stress tensor component $\Pi$.  Using that $\Pi = P_0 - {\cal P}_L$ one obtains
\be
\overline\Pi(\xi) \equiv \frac{\Pi}{\epsilon} = \frac{1}{3} \left[ 1 - \frac{{\cal R}_L(\xi)}{\cal R(\xi)} \right]  .
\label{eq:pixirel}
\ee 
In the left panel of Fig.\ \ref{fig:pibar} we plot $\overline{\Pi}$ as a function of $\xi$ determined via Eq.~(\ref{eq:pixirel}) and, in the right panel, we plot $\xi$ as a function of $\overline{\Pi}$ determined via numerical inversion of Eq.~(\ref{eq:pixirel}).  We note, importantly, that in aHydro $\overline\Pi$ is bounded, $-2/3 < \overline{\Pi} < 1/3$.  This is related to the positivity of the longitudinal and transverse pressures which naturally emerges in this framework. Furthermore, $\overline\Pi$ is related to the inverse Reynolds number \cite{Denicol:2012cn} via
\be
R_\pi^{-1} = \frac{\sqrt{\Pi^{\mu\nu} \Pi_{\mu\nu}}}{P_0} = 3 \sqrt{\frac{3}{2}} |\overline\Pi| \, .
\label{eq:reynoldsnumber}
\ee
As a consequence, a series in $\overline\Pi$ can be roughly understood as an expansion in $R_\pi^{-1}$.

We will also need the relation between the time derivatives of $\Pi$ and $\xi$ which can be obtained from Eq.~(\ref{eq:pixirel})
\be
\frac{\dot\Pi}{\epsilon} = \overline\Pi^\prime \dot\xi + \overline\Pi \partial_\tau \!\log\epsilon \, ,
\ee
which upon using Eqs.~(\ref{eq:pixirel}) and (\ref{eq:firstmom}) gives
\be
\dot\xi = \frac{1}{\overline\Pi'} \left[ \frac{\dot\Pi}{\epsilon} + \frac{\Pi}{\epsilon\tau} \left( \frac{4}{3} - \frac{\Pi}{\epsilon}  \right)  \right] ,
\label{eq:xidot2}
\ee
where $\overline\Pi' \equiv d\overline\Pi/d\xi$.

Plugging (\ref{eq:xidot2}) into (\ref{eq:2ndmomf}), one obtains
\be
\frac{\dot\Pi}{\epsilon} + \frac{\Pi}{\epsilon\tau} \left( \frac{4}{3} - \frac{\Pi}{\epsilon}  \right) - \left[ \frac{2(1+\xi)}{\tau} - \frac{{\cal H}(\xi)}{\tau_{\rm eq}} \right]\overline\Pi'(\xi) = 0\, ,
\label{eq:2ndmomf3}
\ee
with
\be
{\cal H}(\xi) \equiv \xi (1+\xi)^{3/2}{\cal R}^{5/4}(\xi) \, ,
\ee
and the understanding that $\xi = \xi(\overline\Pi)$ with $\xi(\overline\Pi)$ being the inverse function of $\overline\Pi(\xi)$ (shown in the right panel of figure \ref{fig:pibar}).  Written in this form, we can see explicitly that the aHydro second-moment equation sums an infinite number of terms in the expansion in the inverse Reynolds number (\ref{eq:reynoldsnumber}).  In the next section we will expand this equation in powers of the inverse Reynolds number through second order in order to compare it to standard viscous hydrodynamics.

\subsubsection{Small $\xi$ expansion}

In order to make the final connection to standard viscous hydrodynamics, one can expand Eq.~(\ref{eq:2ndmomf3}) in $\xi$ around $\xi=0$.\footnote{The Taylor series around $\xi=0$ has a finite radius of convergence and converges for $|\xi|<1$ due to the cut in the ${\cal H}$ function at $\xi=-1$.}  For this purpose we need the $\xi$ expansions of the various functions that appear in this formalism to construct an explicit inversion and rewrite the equations solely in terms of $\overline\Pi$.  At second-order in $\xi$, one finds
\ba
\overline\Pi &=& \frac{8}{45} \xi  \left[1 - \frac{13}{21} \xi + {\cal O}(\xi^2) \right] , \nonumber \\
\overline\Pi^\prime  &=& \frac{8}{45} \left[1 - \frac{26}{21} \xi + \frac{131}{105} \xi^2 + {\cal O}(\xi^3) \right] , \nonumber \\
(1+\xi)\overline\Pi^\prime  &=& \frac{8}{45} \left[1 - \frac{5}{21} \xi + \frac{1}{105} \xi^2 + {\cal O}(\xi^3) \right] , \nonumber \\
{\cal H} &=& \xi + \frac{2}{3} \xi^2 + {\cal O}(\xi^3) \, .
\ea
Inverting the relationship between $\overline\Pi$ and $\xi$ to second-order in $\overline\Pi$ gives
\be
\xi = \frac{45}{8} \overline\Pi \left[ 1 + \frac{195}{56} \overline\Pi + {\cal O}(\Pi^2)  \right] , 
\ee
which results in
\ba
\overline\Pi^\prime  &=& \frac{8}{45} - \frac{26}{21} \overline\Pi + \frac{1061}{392} \overline\Pi^2 + {\cal O}(\overline\Pi^3) \, , \nonumber \\
(1+\xi)\overline\Pi^\prime  &=& \frac{8}{45} - \frac{5}{21} \overline\Pi - \frac{38}{49} \overline\Pi^2 + {\cal O}(\overline\Pi^3) \, , \nonumber \\
{\cal H} &=& \frac{45}{8} \overline\Pi \left[ 1 + \frac{405}{56} \overline\Pi + {\cal O}(\overline\Pi^3) \right] ,  \nonumber \\
{\cal H} \overline\Pi^\prime &=&  \overline\Pi + \frac{15}{56}  \overline\Pi^2 + {\cal O}(\overline\Pi^3) \, .
\ea

Applying this to the equation of motion (\ref{eq:2ndmomf3}) and keeping only terms through $\Pi^2$ gives
\be
\dot\Pi - \frac{4 \eta}{3 \tau_\pi \tau} + \frac{38}{21} \frac{\Pi}{\tau} - \frac{36\tau_\pi}{245\eta} \frac{\Pi^2}{\tau}
= - \frac{\Pi}{\tau_\pi} - \frac{15}{56} \frac{\Pi^2}{\tau_\pi \epsilon} \, %
\label{eq:2ndmomf4}
\ee
where, on the left hand side, we have used the fact that one can eliminate the energy density by expressing it in terms of the transport coefficients
\be
\epsilon = \frac{15}{4} \frac{\eta}{\tau_{\rm eq}} \, ,
\ee
and relabeled $\tau_{\rm eq} \rightarrow \tau_\pi$ in order cast the equations in ``standard" second order hydrodynamics form.  Note that, to linear order in $\Pi$, Eq.~(\ref{eq:2ndmomf4}) agrees with previously obtained RTA second-order viscous hydrodynamics results \cite{Denicol:2010xn,Denicol:2012cn,Denicol:2014loa,Jaiswal:2013vta,Jaiswal:2013npa}.  However, at order $\Pi^2$, the value of $\lambda_1$ implied is $\lambda_1 = \eta \tau_\pi/7$ which is different by a factor of five compared with prior reported values \cite{Romatschke:2011qp,Ling:PC} which obtained instead $\lambda_1 = 5\eta \tau_\pi/7$.\footnote{The coefficient $\lambda_1$ emerges in the literature because the $\Pi^2$ term appearing on the RHS is traditionally written in the form $\lambda_1 \Pi^2/(2 \tau_\pi \eta^2)$.}  In addition, compared to the standard second-order hydro result, at second order in the $\xi$ expansion we find the appearance of an additional term in the form of the last term on the left-hand side of (\ref{eq:2ndmomf4}). Such term goes beyond the standard truncation order used in the derivation of the DNMR equations \cite{Denicol:2012cn} since it is formally of $\mathcal{O}(K_N R_\pi^{-2})$.


\section{Attractor dynamics in different models}
\label{sec:attractorvars}

In this section we investigate the hydrodynamic attractor behavior of aHydro and compare it with the corresponding results in MIS and DNMR theories. In all of these three cases, the system's dynamics is determined by solving the differential equations for $\epsilon$ and $\Pi$. To make contact with previous studies, however, we follow \cite{Heller:2015dha} and introduce the dimensionless ``time'' variable
\be
w \equiv \tau T(\tau) \, .
\ee
with which one may define the {\em amplitude} 
\be
\varphi(w) \equiv \tau \frac{\dot w}{w} = 1 + \frac{\tau}{4} \partial_\tau\!\log \epsilon \, ,
\label{wdot1}
\ee
which is related to $\Pi$ as follows
\be
\frac{\Pi}{\epsilon} = 4\left(\varphi -\frac{2}{3}\right). 
\ee 
From this we see that a solution for the proper-time evolution of the energy density uniquely specifies the $w$-dependence of the amplitude $\varphi$, as it should be. Also, we note that the positive energy condition \cite{Janik:2005zt} imposes that $\varphi$ is bounded in the region $0 \leq \varphi \leq 1$. 

The change of variables from $\{\epsilon,\Pi\} \to \{w,\varphi\}$ is convenient because it allows one to express the coupled set of first-order ODEs for $\{\epsilon,\Pi\} $ in terms of a single first-order ODE for $\varphi(w)$ \cite{Heller:2015dha}. In the case of MIS and DNMR, this procedure gives
\be
c_\pi w \varphi \varphi' + 4 c_\pi \varphi^2 + \left[ w + \left( \beta_{\pi\pi} - \frac{20}{3} \right) c_\pi \right] \varphi - \frac{4 c_\eta}{9} -\frac{2c_\pi}{3} ( \beta_{\pi\pi} - 4) - \frac{2w}{3}  = 0\,
\label{attractor2ndorder} 
\ee
where $\varphi' = d\varphi(w)/dw$, $c_\pi \equiv \tau_\pi T$, and $c_\eta = \eta/s$ (with $c_\pi = 5 c_\eta$ in the cases considered here). After defining the rescaled variable $\overline{w} = w/c_\pi$ one can see that the equation above becomes
\be
\overline{w} \varphi \varphi' + 4 \varphi^2 + \left[ \overline{w} + \left( \beta_{\pi\pi} - \frac{20}{3} \right)  \right] \varphi - \frac{4 c_{\eta/\pi}}{9} -\frac{2}{3} ( \beta_{\pi\pi} - 4) - \frac{2\overline{w}}{3}  = 0\,,
\label{attractor2ndordernew} 
\ee
which makes it clear that the solution only depends on the ratio $c_{\eta/\pi} \equiv c_\eta/c_\pi = (\eta/s)/(\tau_\pi T)$ and the value chosen for $\beta_{\pi\pi}$. To connect these equations with the RTA Boltzmann one must set $c_{\eta/\pi}=1/5$. Also, we note in passing that $c_{\eta/\pi}$ is the relevant quantity needed in a linearized analysis of the causality and stability properties of MIS-like equations \cite{Hiscock:1983zz,Pu:2009fj}. Using the MIS value $\beta_{\pi\pi}=4/3$ one obtains
\be
\overline{w} \varphi \varphi' + 4 \varphi^2 + \left( \overline{w} - \frac{16}{3} \right) \varphi - \frac{4 c_{\eta/\pi}}{9} + \frac{16}{9} - \frac{2\overline{w}}{3}  = 0 ,
\label{eq:MISeq}
\ee
which agrees precisely with Eq.~(9) of Ref.\ \cite{Heller:2015dha}; however, for RTA this value for $\beta_{\pi\pi}$ is incorrect. Using the correct value for $\beta_{\pi\pi}=38/21$ one obtains the DNMR RTA equation (again neglecting quadratic terms in $\Pi$)
\be
\overline{w} \varphi \varphi' + 4 \varphi^2 + \left(\overline{w} - \frac{34}{7} \right) \varphi - \frac{4 c_{\eta/\pi}}{9} + \frac{92}{63} - \frac{2\overline{w}}{3}  = 0 \, .
\label{eq:DNMReq}
\ee
Also, we note that, as demonstrated in Eq.~(\ref{eq:2ndmomf4}), aHydro naturally reproduces this equation when truncated at leading order in $\xi$ (linear order in the inverse Reynolds number). 

Following \cite{Heller:2015dha}, attractor behavior can be inferred from Eq.\ \eqref{attractor2ndorder} using a procedure equivalent of the ``slow-roll" expansion in cosmology \cite{Liddle:1994dx}, which in this context may be described as follows. First, one formally introduces a small parameter $\delta$ as a prefactor in the term $ \overline{w} \varphi \varphi'$ in \eqref{attractor2ndordernew} and assume that the solution of the differential equation $\varphi(\overline{w};\delta)$ can be written as power series expansion $\varphi(\overline{w};\delta) = \varphi_0(\overline{w})+\varphi_1(\overline{w})\,\delta + \mathcal{O}(\delta^2)$. After taking into account all orders, one may take the limit $\delta \to 1$. The 0th order truncation is obtained by solving the simple quadratic equation
\be
4  \varphi_0^2 + \left[ \overline{w} + \left( \beta_{\pi\pi} - \frac{20}{3} \right) \right] \varphi_0 - \frac{4 c_{\eta/\pi}}{9} -\frac{2}{3} ( \beta_{\pi\pi} - 4) - \frac{2\overline{w}}{3}  = 0\,
\label{attractor2ndorder0order} 
\ee
and, out of the two possible solutions, the one that is stable and remains finite in the large ``time" (large $\overline{w}$) limit is
\be
\varphi_0(\overline{w}) = \frac{1}{24} \left(-3 \beta_{\pi\pi} +\sqrt{64 c_{\eta/\pi}+(3
   \beta_{\pi\pi} +3 \overline{w}-4)^2}-3 \overline{w}+20\right).
   \label{attractor2ndordersol}
\ee
Though one may easily compute the higher order corrections, in practice one finds that the 0th order solution already represents a good approximation to the exact solution of the differential equation for $\overline{w} > 4$ and $c_{\eta/\pi}=1/5$. 

In this paper we also define an \emph{attractor solution} using the boundary condition  \mbox{$\lim_{\overline{w} \rightarrow 0} \overline{w} \varphi \varphi' = 0$} \cite{Heller:2015dha}, which then implies that
\be
\lim_{\overline{w}\to 0}\varphi(\overline{w})=\frac{1}{24} \left(-3 \beta_{\pi\pi} +\sqrt{64 c_{\eta/\pi}+(3
   \beta_{\pi\pi} -4)^2}+20\right).
\ee 
This gives a smooth curve that necessarily agrees with the 0th order solution at $\overline{w}=0$ and also at late times.   In the next section we generalize the analysis performed here to determine the attractor dynamics of aHydro.

\subsection{aHydro attractor}

In this section we present our final dynamical equation for aHydro after recasting the two first-order differential equations as a single second-order differential equation written in terms of $\varphi$ and $w$. In order to obtain the aHydro dynamical equation, we must combine the following identity
\be
w \varphi \varphi' = -\frac{8}{3} + \frac{20}{3}\varphi - 4\varphi^2 + \frac{\tau}{4}\frac{\dot\Pi}{\epsilon}
\label{eq:finalfirstmom}
\ee
and \eqref{eq:2ndmomf3}.  To do this we should first express  Eq.~(\ref{eq:2ndmomf3}) in terms of $\varphi$ and $w$.  Using that $\tau\partial_\tau\!\log \epsilon = 4(\varphi - 1)=-4/3+\Pi/\epsilon$, one finds from Eq.~(\ref{eq:2ndmomf3}) 
\be
\frac{\tau}{4} \frac{\dot\Pi}{\epsilon} =  \frac{8}{3} - \frac{20}{3} \varphi + 4 \varphi^2 + \left[ \frac{1}{2} (1+\xi)  - \frac{w}{4 c_\pi} {\cal H} \right] \overline\Pi' \,.
\ee
Plugging this into Eq.~(\ref{eq:finalfirstmom}) gives our final result for the aHydro attractor equation
\be
\overline{w} { \varphi} \frac{\partial \varphi}{\partial \overline{w}}  = \left[ \frac{1}{2} (1+\xi) - \frac{\overline{w}}{4} {\cal H} \right] \overline\Pi' \, .
\label{eq:ahydroattractoreq2}
\ee
Note that above $\xi = \xi(\overline\Pi) = \xi(4\varphi - 8/3)$ and likewise for $\overline\Pi'$.  We remark that the aHydro equation derived above does not depend explicitly on $c_{\eta/\pi}$ - the aHydro attractor solution is universal if plotted as a function of $\overline{w}$. Since we work in relaxation-time approximation, this is true for the other second order hydrodynamic approximations (i.e., $c_{\eta/\pi}$ must be set to be $1/5$ for RTA dynamics) presented above as well. 

The aHydro equation \eqref{eq:ahydroattractoreq2} the gradient expansion series solution in powers of $1/\overline{w}$ has zero radius of convergence \cite{Florkowski:2016zsi}. Thus, the solution of the differential equation \eqref{eq:ahydroattractoreq2} may also be considered to be a resummation of the gradient series, as in MIS theory \cite{Heller:2015dha}. However, we emphasize that the right-hand-side of Eq.~(\ref{eq:ahydroattractoreq2}) also includes a sum of an infinite number terms in the inverse Reynolds number, which is conceptually different than DNMR which derived their equations of motion assuming a perturbative series in $R_\pi^{-1}$.

In the case of aHydro, even the 0th order approximation in the slow-roll expansion must be solved numerically so we skip directly to the solution of the differential equation.  Again, for this purpose, the attractor solution is obtained by imposing the same boundary condition as before at $\overline{w}=0$.  Using the numerical solution of the approximate equation, one finds
\be
\lim_{\overline{w} \rightarrow 0} \varphi(\overline{w}) =  \frac{3}{4} \, .
\label{eq:ahydrobc}
\ee
With this we simply numerically solve Eq.~(\ref{eq:ahydroattractoreq2}).  Note that the limit above guarantees the positivity of the longitudinal pressure of the attractor solution at all points in the plasma as $\overline{w} \rightarrow 0$.

\subsection{Exact RTA attractor solution}

In addition to comparing the attractors emerging from different hydrodynamic theories, we will also determine the attractor which emerges from exact solution of the RTA Boltzmann equation. For this case, one can write down an integral equation which can be numerically solved to arbitrary accuracy \cite{Florkowski:2013lza,Florkowski:2013lya}
\be
\bar{\cal E}(\tau) = D(\tau,\tau_0) \,
\frac{{\cal R}\big(\xi_{\rm FS}(\tau)\big)}{{\cal R}\left(\xi_0\right)}
+ \int_{\tau_0}^{\tau} \! \frac{d\tau^\prime}{\tau_{\rm eq}(\tau^\prime)} \, D(\tau,\tau^\prime) \, 
\bar{\cal E}(\tau^\prime) \, {\cal R}\!\left( \! \left(\frac{\tau}{\tau^\prime}\right)^2 - 1 \right) ,
\label{eq:inteq}
\ee
where ${\bar{\cal E} = {\cal E}/{\cal E}_0}$ is the energy density scaled by the initial energy density, ${\cal R}$ is defined in Eq.~(\ref{eq:rfunc}),  $\xi_0$ is the initial momentum-space anisotropy, ${\xi_{\rm FS}(\tau) = (1+\xi_0)(\tau/\tau_0)^2-1}$, and
\be
{D(\tau_2,\tau_1) = \exp\!\left[-\int_{\tau_1}^{\tau_2} d\tau^{\prime\prime} \, \tau^{-1}_{\rm eq}(\tau^{\prime\prime})\right]} ,
\ee
is the damping function.  A procedure for obtaining the attractor from this integral equation is explained in Ref.~\cite{Romatschke:2017vte}. However, in practice it amounts to using an infinitely oblate anisotropic initial condition $\xi_0 \rightarrow \infty$ as the solution to this integral equation and taking the initial proper time arbitrarily small.  A C-code for solving this integral equation can be downloaded using the URL specified in Ref.~\cite{MikeCodeDB}.

\section{Results and discussion}
\label{results}

\begin{figure*}[t!]
\centerline{
\includegraphics[width=.48\linewidth]{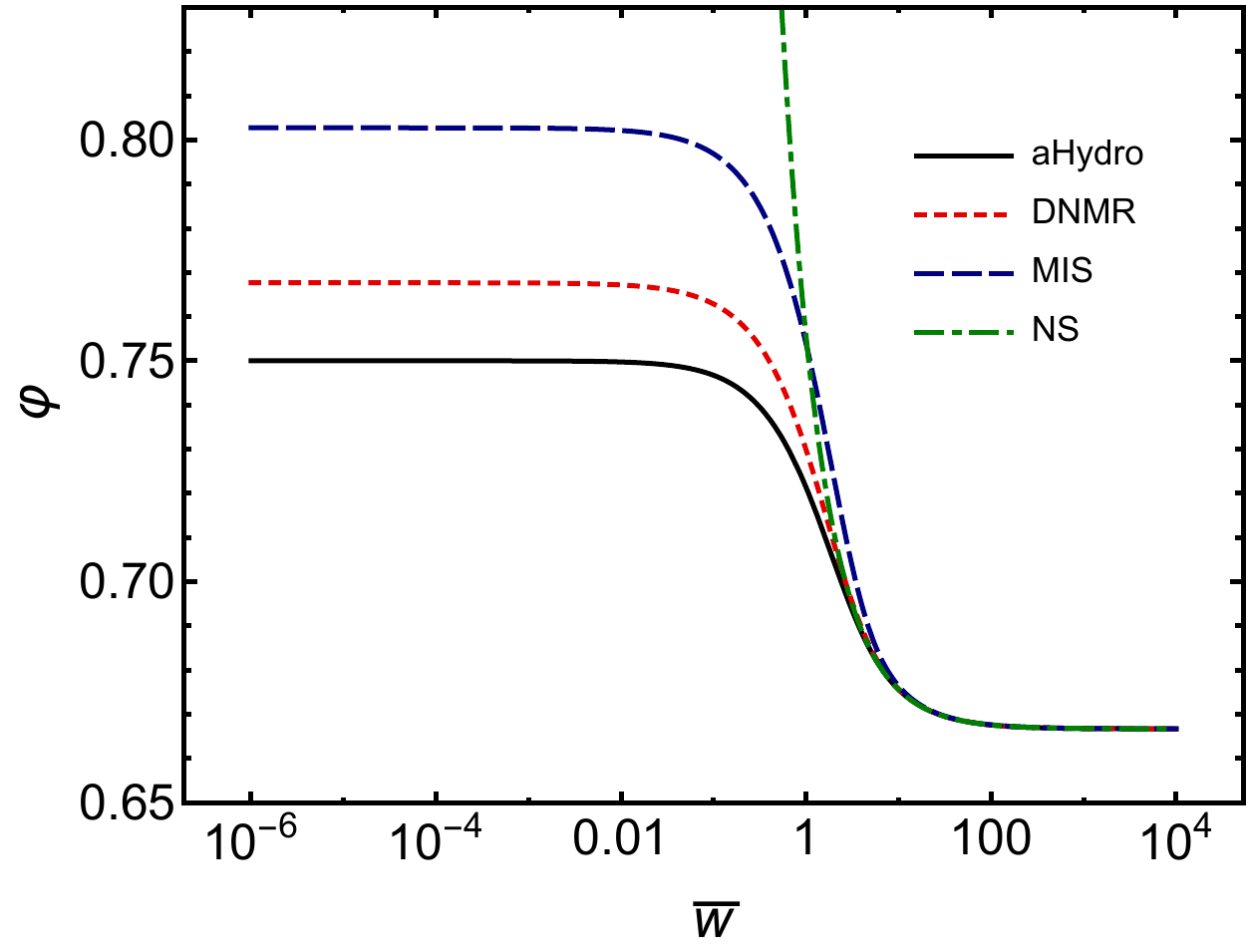}
\hspace{3mm}
\includegraphics[width=.48\linewidth]{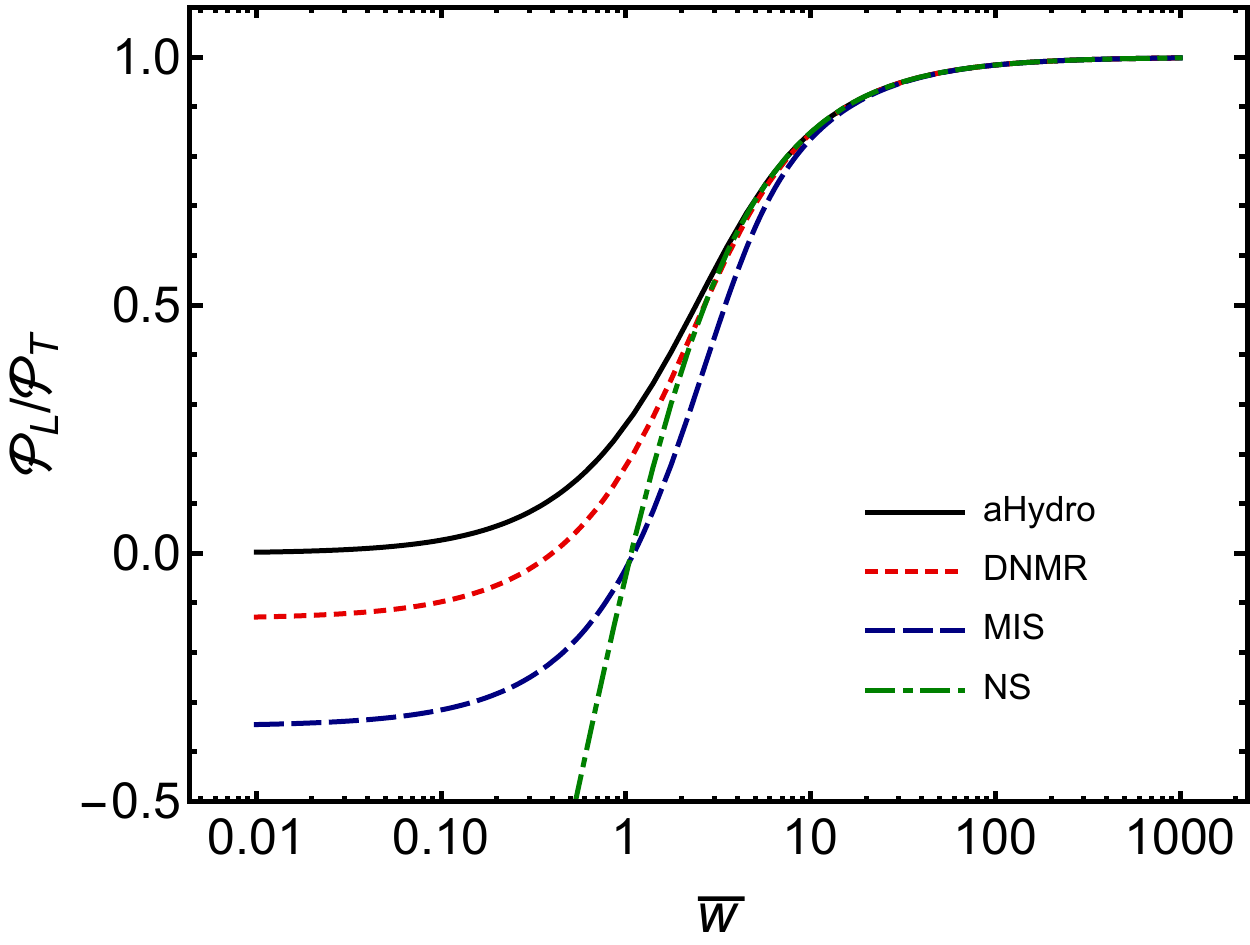}
}
\caption{(Color online) The left panel shows the solution for $\varphi$ and the right panel shows the solution for the corresponding pressure ratio ${\cal P}_L/{\cal P}_T$. }
\label{fig:attractor_2}
\end{figure*}

\begin{figure*}[t!]
\centerline{
\includegraphics[width=.5\linewidth]{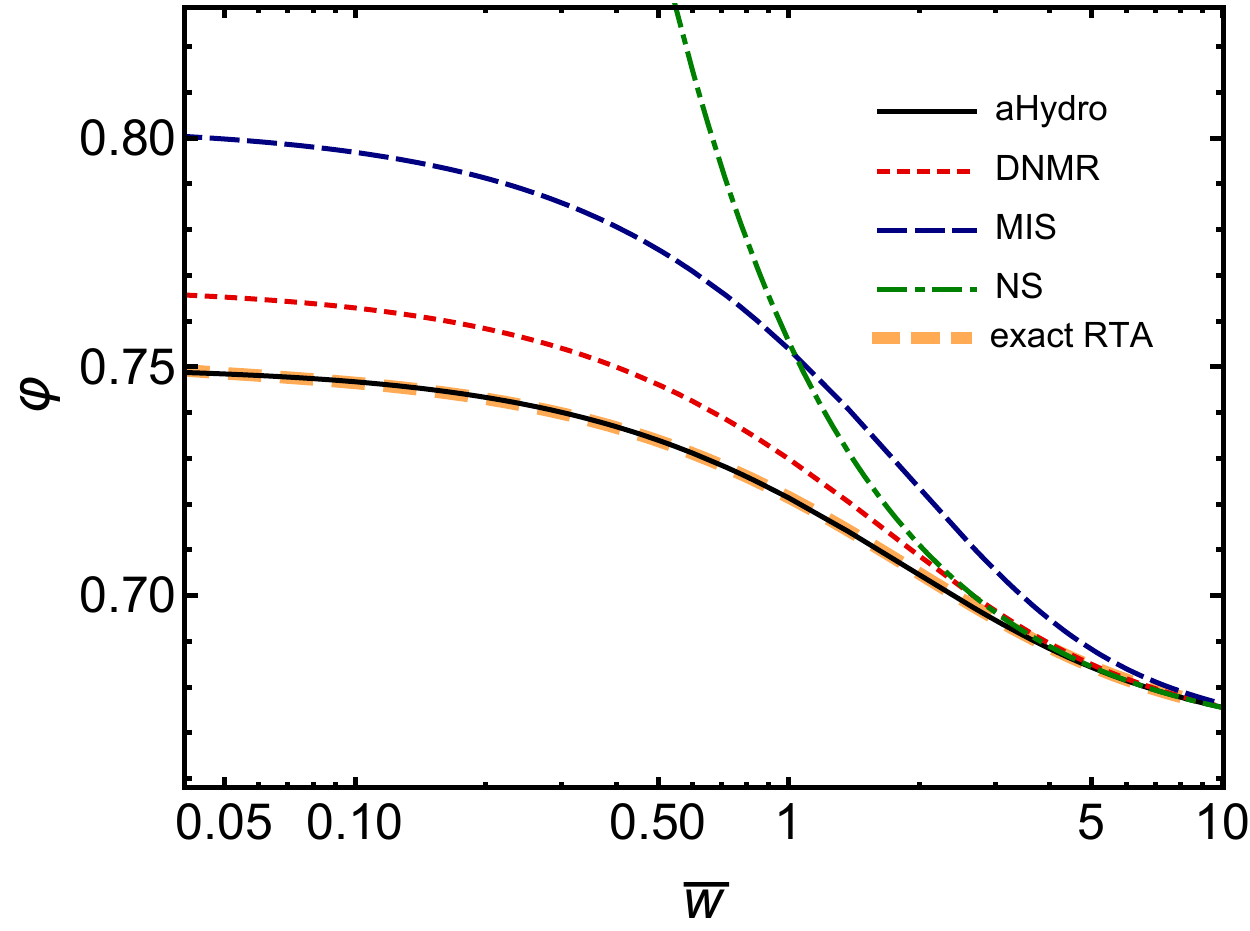}
}
\caption{(Color online) aHydro, MIS, and DNMR attractors compared to the attractor obtained from exact solution to the RTA Boltzmann equation.}
\label{fig:attractor_2_compare}
\end{figure*}

In Fig.\ \ref{fig:attractor_2} we compare the attractors for $\varphi(w)$ determined using the solution of the differential equation in each case in the left panel, i.e. Eqs.~(\ref{eq:MISeq}), (\ref{eq:DNMReq}), and (\ref{eq:ahydroattractoreq2}), subject to their corresponding boundary conditions at $\overline{w}=0$ mentioned in the last section.  In the right panel we show the corresponding longitudinal to transverse pressure ratio which can be computed using
\be
\frac{{\cal P}_L}{{\cal P}_T} = \frac{3 - 4 \varphi}{2 \varphi - 1} \, .
\ee
Using the criteria that ${\cal P}_L/{\cal P}_T > 0.9$, we observe that approximate isotropization only occurs for $\overline{w} > 15$. Also, we note that, depending on the differential equation used to determine the attractor solution, $\varphi$ might exceed $3/4$, which will cause this ratio to go negative due to a negative longitudinal pressure.  As can be seen from the right panel, both the MIS and DNMR attractors ``pull'' the system towards negative longitudinal pressures since $\varphi > 3/4$ at early times corresponding to small $\overline{w}$.  This behavior does not occur in aHydro since, in this case, $1/2 < \varphi < 3/4$. 

Next we turn to Fig.\ \ref{fig:attractor_2_compare} where we compare the aHydro, MIS, and DNMR attractors to the corresponding quantity obtained from the exact solution to the 0+1d RTA Boltzmann equation (\ref{eq:inteq}).  Additionally, in Fig.~\ref{fig:attractor_2_compare} we include a curve showing the Navier-Stokes (NS) result~\cite{Heller:2015dha}
\be
\varphi_{\rm NS} = \frac{2}{3} + \frac{4}{9} \frac{c_{\eta/\pi}}{\overline{w}} \, .
\ee
which can be obtained by taking the $w \rightarrow \infty$ limit of \eqref{attractor2ndordersol} and truncating at the first non-trivial order.  As Fig.~\ref{fig:attractor_2_compare} demonstrates, the aHydro attractor solution is virtually indistinguishable from the exact RTA attractor.  In fact, it is unclear to us whether the remaining differences, being maximum of 0.04\% in the range shown, might be purely numerical in origin. Since aHydro involves not only a resummation in Knudsen number but also in the inverse Reynolds number, the excellent agreement found between the aHydro solution and the exact kinetic theory result suggests that the inverse Reynolds number resummation may also be a property of the latter. This may serve as a guide to derive other new approaches to far-from-equilibrium hydrodynamics that do not rely on a perturbative treatment of both the Knudsen and the inverse Reynolds number series, which may be particularly useful in the search for a novel (causal and stable) hydrodynamic theory that incorporates the quasinormal oscillatory behavior found at strong coupling using holography \cite{Denicol:2011fa,Noronha:2011fi,Heller:2014wfa}.   

Turning to the second order approaches, we see that the DNMR attractor is in significantly better agreement with the exact RTA attractor solution than MIS, as one might expect since the MIS equations have the incorrect value of $\beta_{\pi\pi}$ within RTA.  In this plot, the NS solution is included to emphasize that this approximation, although previously thought of as the late-time attractor, does not coincide with the attractor solution until one reaches quite large values of $\overline{w}$ (i.e., sufficiently close to local equilibrium). 

\begin{figure*}[t!]
\centerline{
\includegraphics[width=.48\linewidth]{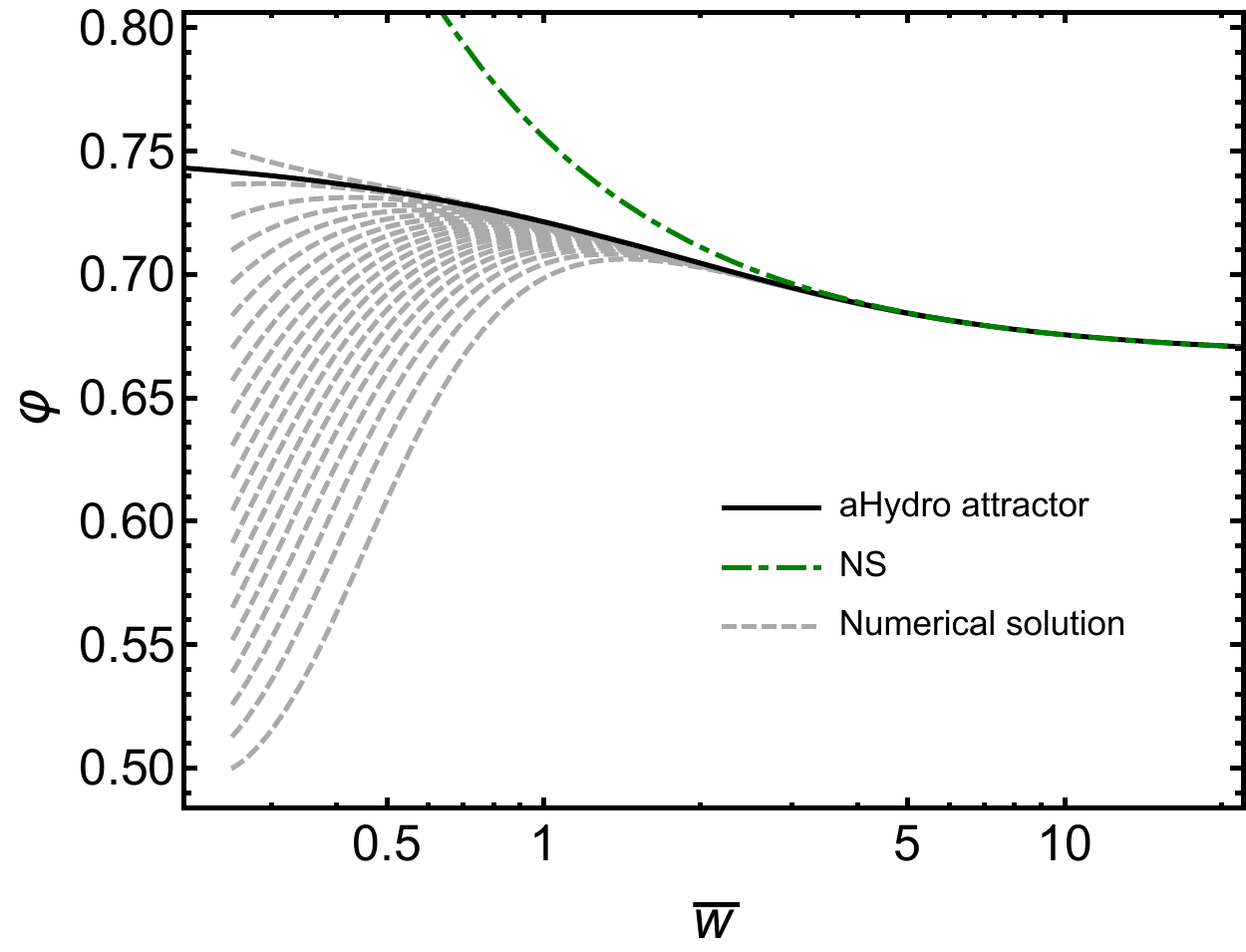}
\hspace{3mm}
\includegraphics[width=.48\linewidth]{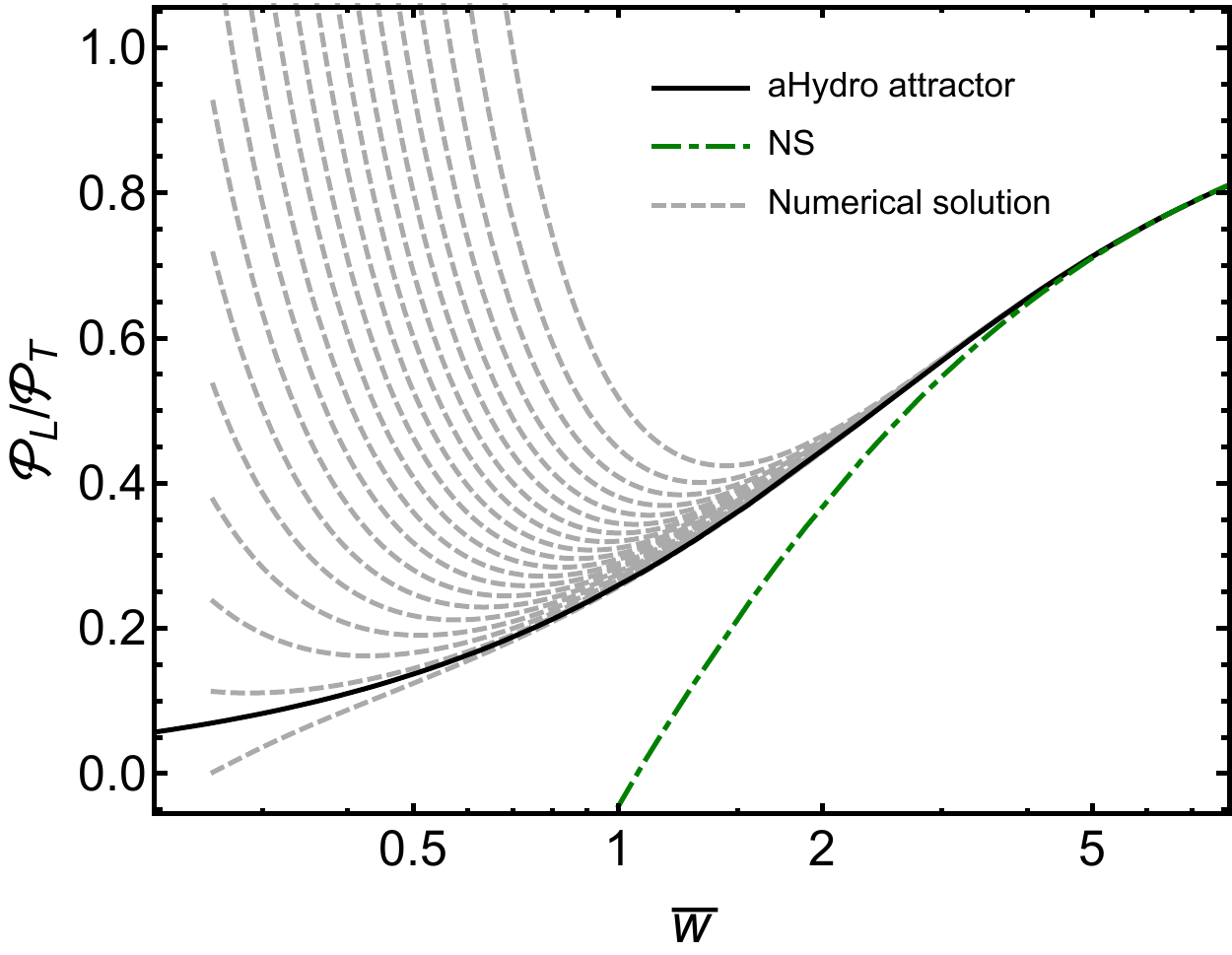}
}
\caption{(Color online) aHydro attractor (solid black line) and numerical solutions (grey dashed lines) corresponding to a variety of initial conditions for $\Pi$.  The left panel shows the solution for $\varphi$ and the right panel shows the solution for the corresponding pressure ratio ${\cal P}_L/{\cal P}_T$. }
\label{fig:ahydro_attractor_2}
\end{figure*}

\begin{figure*}[t!]
\centerline{
\includegraphics[width=.48\linewidth]{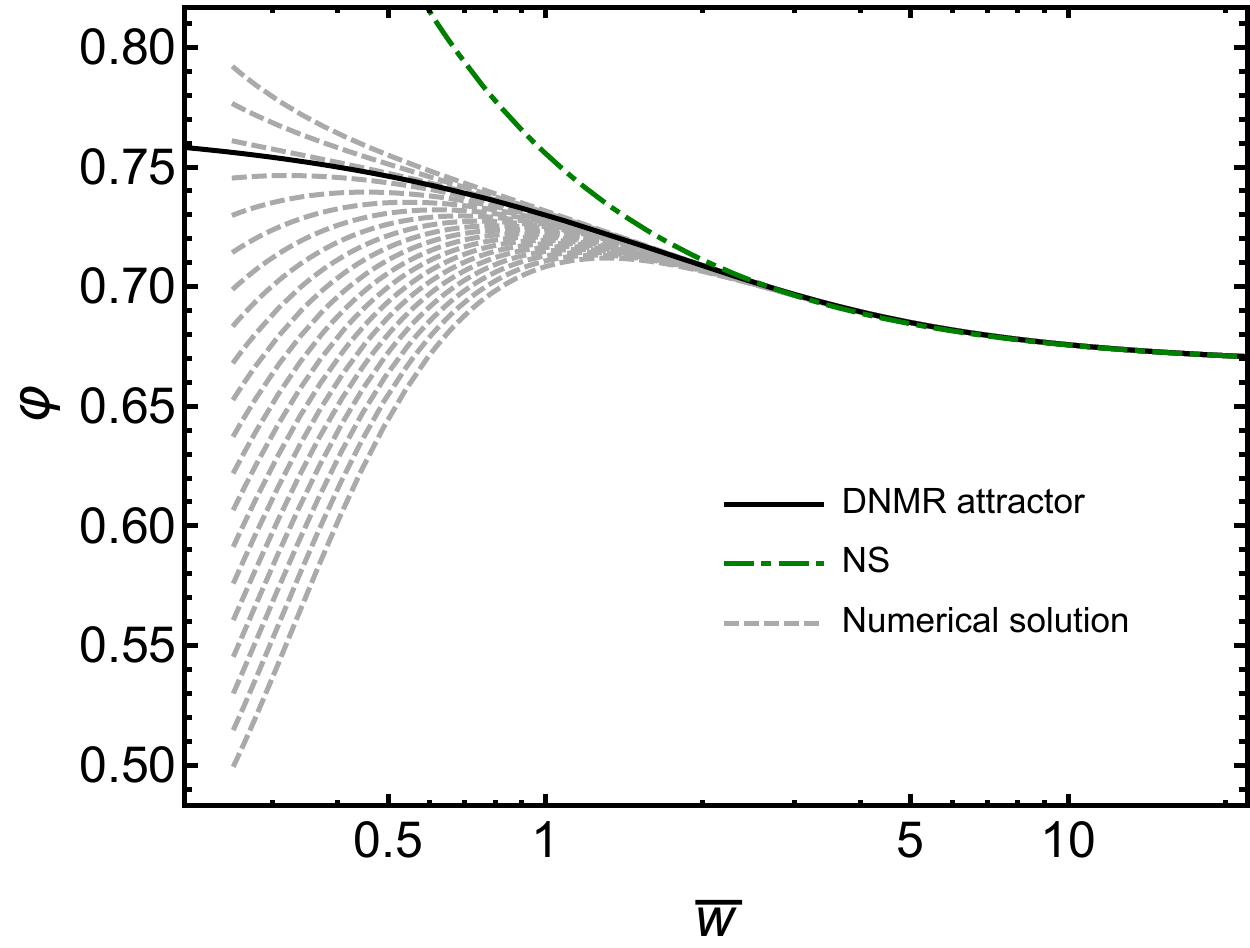}
\hspace{3mm}
\includegraphics[width=.48\linewidth]{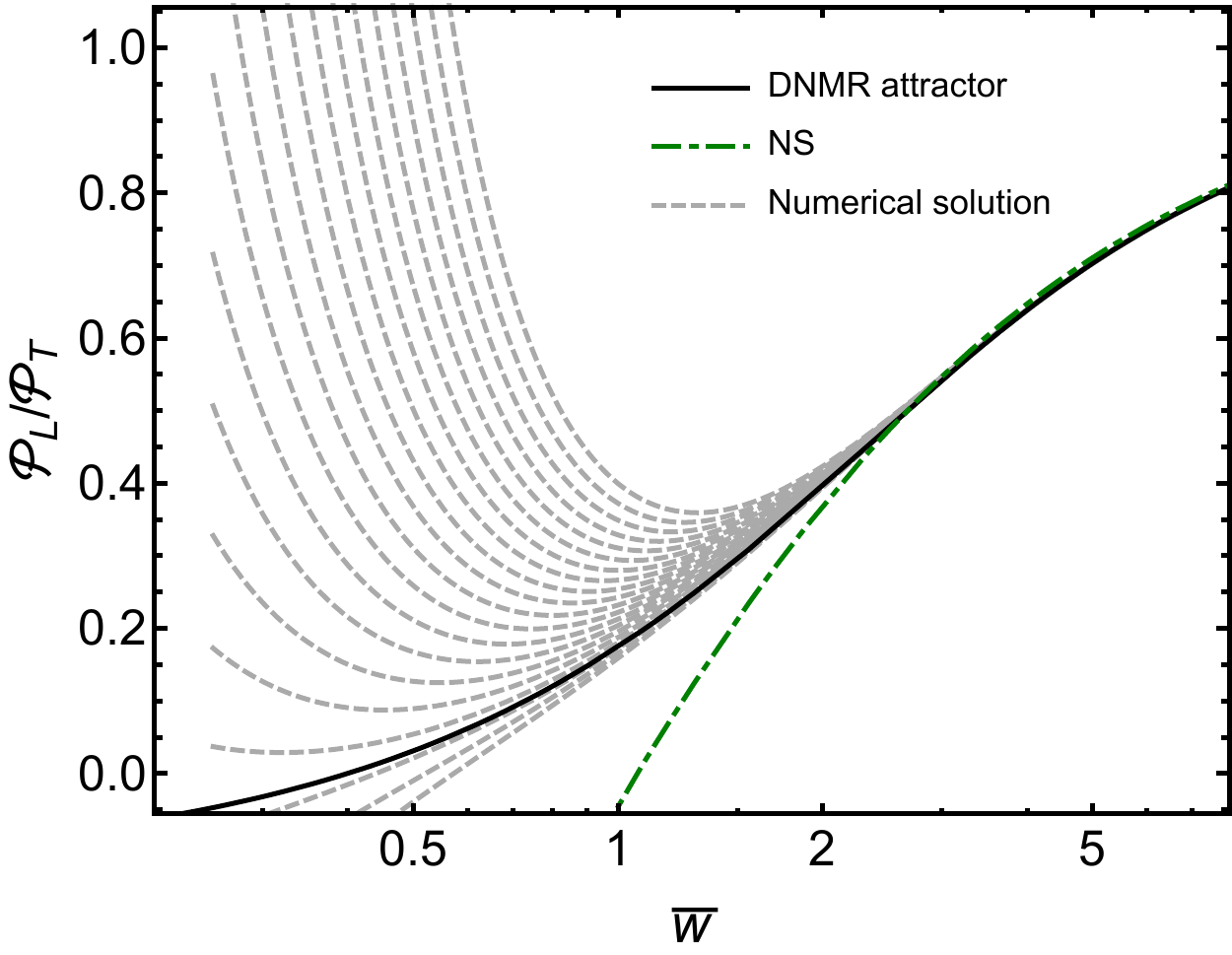}
}
\caption{(Color online) DNMR attractor (solid black line) and numerical solutions (grey dashed lines) corresponding to a variety of initial conditions for $\Pi$.  The left panel shows the solution for $\varphi$ and the right panel shows the solution for the corresponding pressure ratio ${\cal P}_L/{\cal P}_T$. }
\label{fig:dnmr_attractor_2}
\end{figure*}

Finally, we turn to Figs.\ \ref{fig:ahydro_attractor_2} and \ref{fig:dnmr_attractor_2}.  In these figures we compare the numerical solution of the aHydro and DNMR dynamical equations along with their respective attractors and the NS solution.  For the numerical solutions (grey dashed), we fixed an initial energy density $\epsilon_0$ at proper time $\tau_0$ and then varied the initial condition for $\Pi_0$ over a given range.  For both aHydro and DNMR, the numerical solutions shown converge to the attractor solution after approximately \mbox{$\overline{w}_{\rm attractor} \sim 2$}. In the context of heavy ion phenomenology, for LHC initial conditions with a central temperature of $T_0 \sim 500$ MeV at $\tau_0 = 0.25$ fm/c and $\eta/s \sim 0.2$, this translates into $\tau_{\rm attractor} \sim 1.3$ fm/c in the center of the fireball. Prior to this time, the system is subject to the evolution of non-hydrodynamic modes and the precise evolution of these modes depends on the microscopic theory under consideration.  Comparing to the NS solution, one reaches the remarkable conclusion that the NS solution is a good approximation quickly after that.  For aHydro and the exact RTA solution, NS starts to be an accurate approximation at $\overline{w}_{\rm NS} \sim 3$ and, for DNMR already at $\overline{w}_{\rm NS}  \sim 2$.  In these examples, we are led to conclude that $\overline{w}_{\rm attractor} \lesssim \overline{w}_{\rm NS}$. For the example at hand one would find \mbox{$\tau_{\rm NS} \sim$ 2.3 fm/c}, which is quite soon after the attractor-driven dynamics kicks in.  However, as we approach the transverse edge of the fireball, the corresponding time scales grow, as does their absolute separation, e.g. in a region with $T_0 \sim 250$ MeV we find $\tau_{\rm attractor} \sim$ 3.4 fm/c and \mbox{$\tau_{\rm NS} \sim$ 6 fm/c} assuming, again, that $\eta/s=0.2$ and is constant.  If $\eta/s$ increases at low temperatures these time scales would increase proportionally.  Applying this as a rough guide for full 3+1d simulations, one would conclude that low-temperature regions of the plasma (e.g. the edges) would still be particularly sensitive to non-hydrodynamic modes.  

\section{Conclusions and Outlook}
\label{conclusions}

In this paper we obtained the dynamical attractors associated with the aHydro and DNMR versions of viscous hydrodynamics.  Along the way we demonstrated that the aHydro dynamical equations resum an infinite number of terms in the inverse Reynolds number, which does not occur in other approaches.  As a direct consequence of this all-order resummation, we found that (a) the resulting aHydro attractor was naturally restricted to $1/2 < \varphi < 3/4$ which guarantees the positivity of both the longitudinal and transverse pressures and (b) the resulting aHydro attractor was virtually indistinguishable from the attractor emerging from exact solution of the RTA Boltzmann equation.  On the DNMR front, we demonstrated that it provides a significant improvement over the MIS attractor when compared to the exact RTA solution due to the systematic inclusion of all second-order contributions (taken into account in the coefficient $\beta_{\pi\pi}$).  We also showed that, when truncated at leading order in the inverse Reynolds number, the aHydro dynamical equations identically reduce to the DNMR equations.

As part of the results presented we compared the numerical solution of the aHydro and DNMR equations with their respective attractor solutions and found that, similar to other frameworks, the numerical solutions for a variety of different initial conditions approach the attractor solution within a time $\tau_{\rm attractor}$.  In LHC heavy-ion collisions, one expects initial temperatures \mbox{$T_0 \lesssim$ 500 MeV} at $\tau_0 = 0.25$ fm/c and $\eta/s \sim 0.2$, which translates into $\tau_{\rm attractor} \gtrsim $ 1.3 fm/c with the lower bound holding in the hot center of the fireball on average.  Prior to $\tau \sim \tau_{\rm attractor}$, each local region of the system is subject to the evolution of non-hydrodynamic modes \cite{Noronha:2011fi,Heller:2014wfa,Bazow:2015zca,Florkowski:2017olj} whose precise evolution depends on the microscopic theory being considered and whose ``lifetime'' increases as one approaches the low-temperature edge of the plasma.  As such, the dynamics of the system prior to $\tau_{\rm attractor}$ is non-universal.  

Faced with such a situation it becomes critically important to identify the appropriate microscopic theory to describe the dynamics of the system.  In the center of the fireball, where the energy densities are the largest at early times, one would expect approaches that interpolate between perturbative QCD and holography to be the most appropriate. However, as one approaches the dilute edges a formulation in terms of hadronic kinetic theory would seem to be the most appropriate.  Since some of these regions could, in principle, be described in terms of the Boltzmann or Boltzmann-Vlasov equations and the same theories match smoothly onto the late-time hydrodynamical attractor,  this motivates the ongoing study of hydrodynamic theories that can be obtained from relativistic kinetic theory. Further progress may be obtained once more realistic nonlinear collision kernels are included to investigate the properties of the kinetic theory attractor, such as in \cite{Kurkela:2015qoa} and \cite{Bazow:2015dha,Bazow:2016oky}, where the microscopic dynamics is much more complex than the single relaxation timescale used in the relaxation time approximation of the Boltzmann equation.

\acknowledgments

M.\ Strickland and J.\ Noronha thank the organizers of the ``Canterbury Tales of Hot QFTs in the LHC Era'' workshop held at St. John's College, Oxford, UK (July 10-14, 2017) for support during the initial stages of this project.  M.\ Strickland was supported by the U.S. Department of Energy, Office of Science, Office of Nuclear Physics under Award No.~DE-SC0013470. J.\ Noronha thanks S\~{a}o Paulo Research Foundation (FAPESP) and Brazilian National Council for Scientific and Technological Development (CNPq) for support and the Department of Physics and Astronomy at Rutgers University for the hospitality. G.\ S.\ Denicol thanks CNPq for support.

\vspace{5mm}

\hrule

\appendix 

\section{An alternative expansion based on the aHydro second-moment method}
\label{appendix1}

\begin{figure*}[t!]
\centerline{
\includegraphics[width=.48\linewidth]{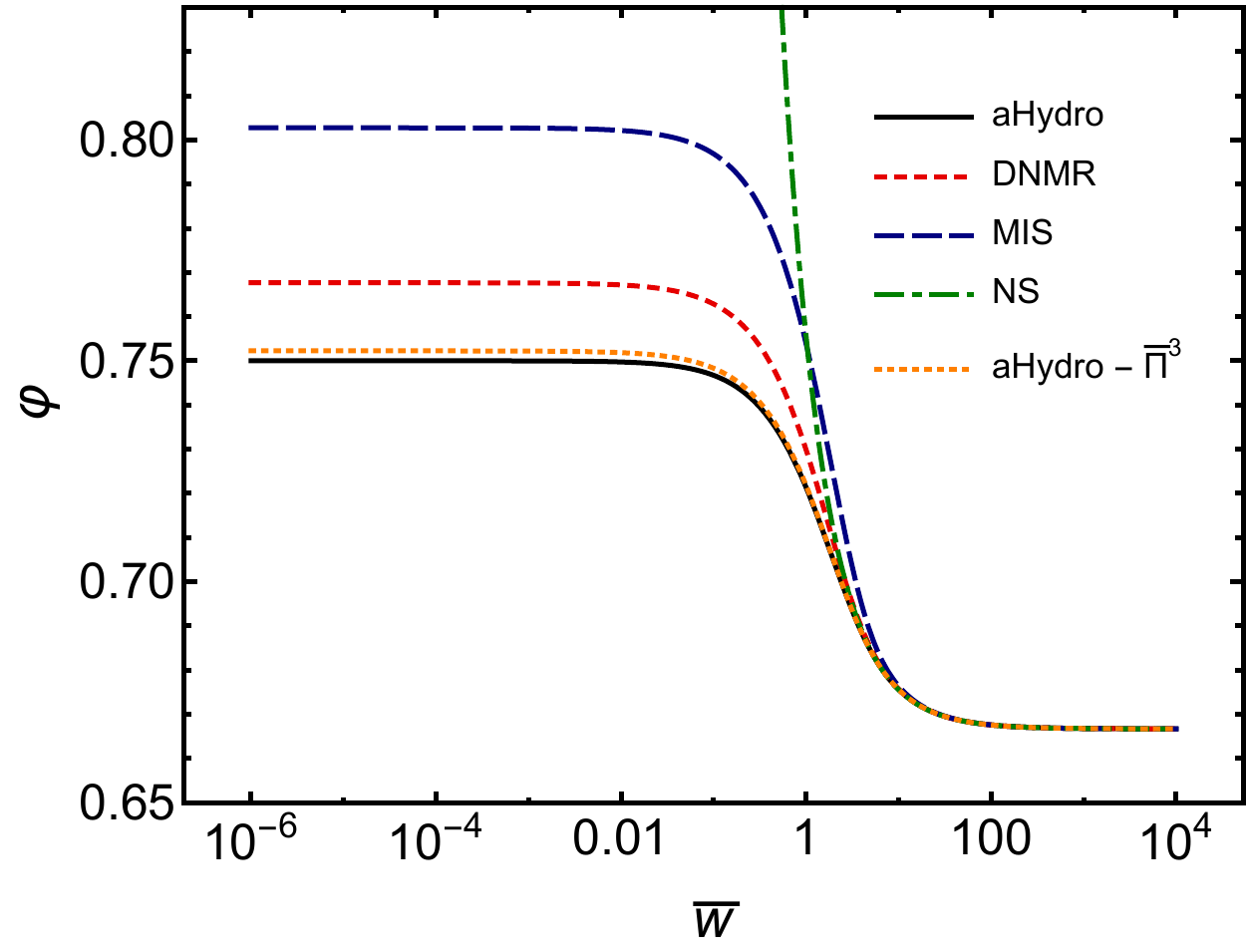}
\hspace{3mm}
\includegraphics[width=.48\linewidth]{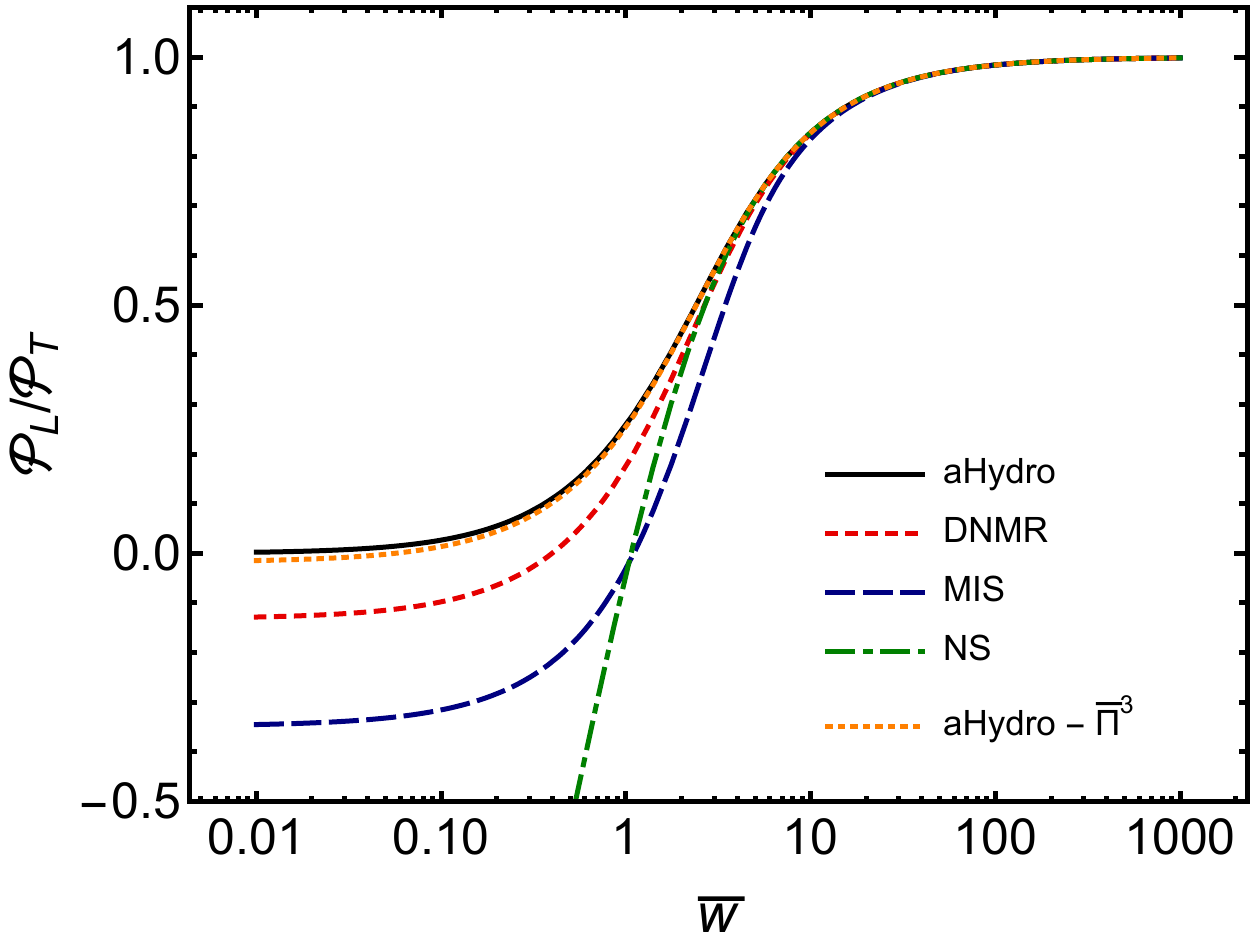}
}
\caption{(Color online) Third-order aHydro attractor (green dot-dashed) compared to the other solutions obtained and presented in the main body of the paper.  The left panel shows the solution for $\varphi$ and the right panel shows the solution for the corresponding pressure ratio ${\cal P}_L/{\cal P}_T$. }
\label{fig:ahydro_attractor_3}
\end{figure*}

In this appendix we consider what happens if we expand Eq.~(\ref{eq:ahydroattractoreq2}) to higher order in $\overline\Pi$ (inverse Reynolds number).  Through order $\overline\Pi^3$, in RTA, one obtains:
\ba
&& \overline{w} { \varphi} \frac{\partial \varphi}{\partial \overline{w}} + \frac{20352 c_{\eta/\pi}}{3773}-\frac{81 \overline{w}}{49} +\left(\frac{603 \overline{w}}{98}-\frac{39070 c_{\eta/\pi}}{3773}\right) \varphi  \nonumber \\
&& \hspace{2cm}
-\left(\frac{44960 c_{\eta/\pi}}{3773}+\frac{1725 \overline{w}}{196}\right)\varphi ^2  + \left(\frac{81000 c_{\eta }}{3773}+\frac{1935 w}{392}\right)\varphi ^3     \, ,
\label{eq:ahydroattractoreqT1}
\ea
with $c_{\eta/\pi} = 1/5$.  The boundary condition necessary is
\ba
\lim_{w \rightarrow 0} \varphi(w) = \frac{1124}{6075}+\frac{7}{12150} \sum_{\sigma=\pm 1} \sqrt[3]{-390484556 + 13365 i \sqrt{281726265}\sigma }  \; \simeq \; 0.752251 \, .
\ea
In Fig.~\ref{fig:ahydro_attractor_3} we plot the solution to the differential equation (\ref{eq:ahydroattractoreqT1}) subject to the above boundary condition.  As can be seen from this figure the third order expansion in $\overline\Pi$ provides a very good approximation of the aHydro attractor.  This expansion can naturally be systematically extended to higher orders.

\section{aHydro attractor using the anisotropic matching principle}
\label{tinti}

Recently, Tinti introduced an alternative method for obtaining the aHydro evolution equations which is based on the so-called ``anisotropic matching principle'' \cite{Tinti:2015xwa}.  In practice, in addition to the equations resulting from the first moment of the Boltzmann equation, following \cite{Denicol:2010xn} one computes the exact equation obeyed by the viscous tensor, plugging in the anisotropic distribution form on the right hand side.  The resulting equation for the pressure difference for a 0+1d conformal system is \cite{Florkowski:2017olj}
\be
\dot\Delta = -\frac{\Delta}{\tau_{\rm eq}} + 2 (1+\xi) \frac{\partial \Delta}{\partial \xi} \, ,
\ee
where 
\be
\Delta \equiv {\cal P}_L - {\cal P}_T = {\cal R}_\Delta(\xi) \epsilon_0(\lambda) = -\frac{3}{2} \Pi \, ,
\label{eq:deltadef}
\ee
and ${\cal R}_\Delta \equiv \left[ {\cal R}_L(\xi) - {\cal R}_T(\xi) \right]/3$.

\begin{figure*}[t!]
\centerline{
\includegraphics[width=.5\linewidth]{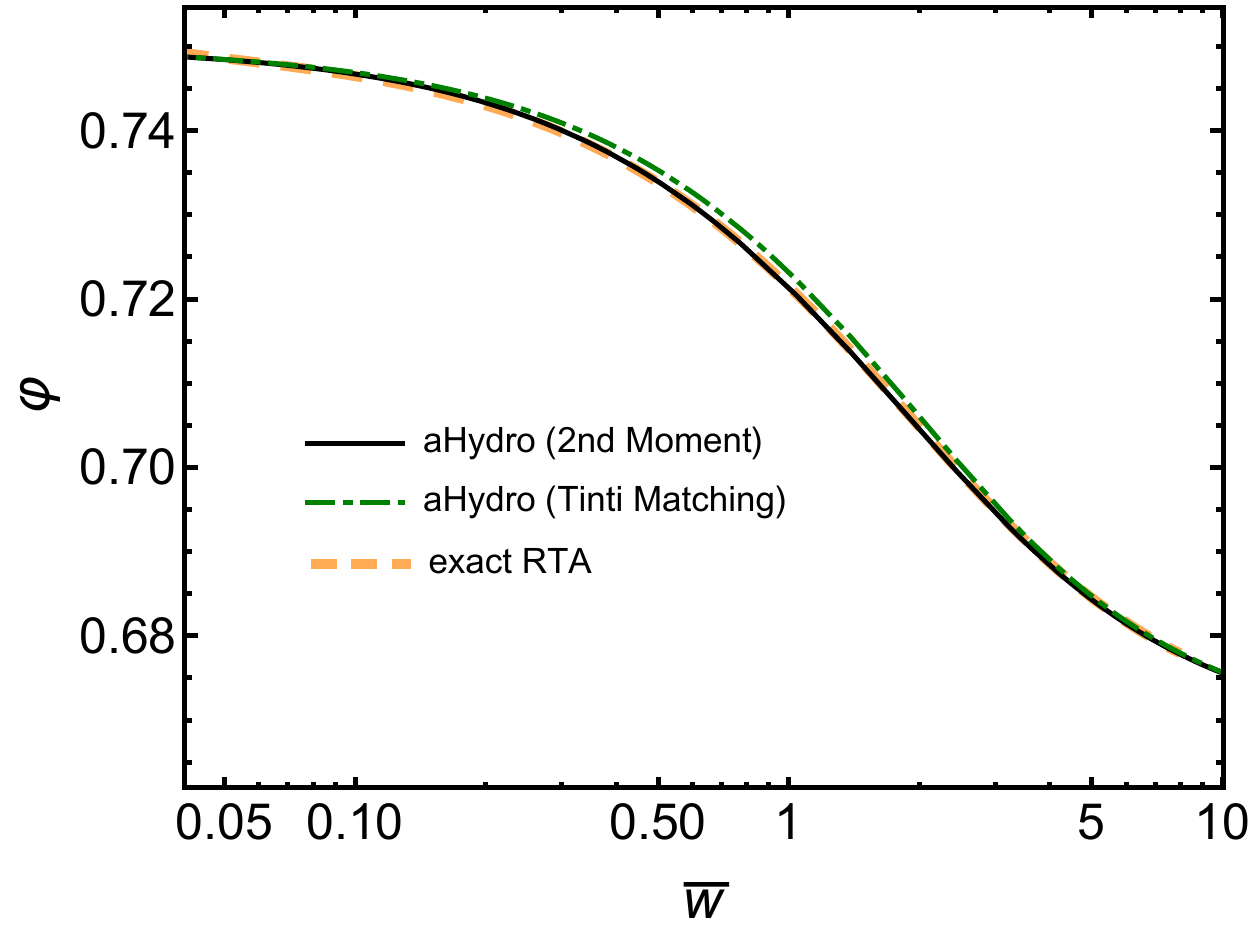}
}
\caption{(Color online) aHydro second-moment based and Tinti's ``anisotropic matching principle'' attractors compared to the attractor obtained from exact solution to the RTA Boltzmann equation.}
\label{fig:tinti_attractor}
\end{figure*}

Using the last equality in Eq.~(\ref{eq:deltadef}), we can write this as an equation for $\Pi$
\be
\frac{\dot\Pi}{\epsilon} = - \frac{\overline\Pi}{\tau_{\rm eq}} - \frac{4}{3} \frac{1+\xi}{\tau} \frac{{\cal R}^\prime_\Delta(\xi)}{{\cal R}(\xi)}.
\ee
Combining this with Eq.~(\ref{eq:finalfirstmom}), one obtains
\be
\overline{w} { \varphi} \frac{\partial \varphi}{\partial \overline{w}}  = -\frac{8}{3} + \frac{20}{3} \varphi - 4 \varphi^2 +\overline{w} \left(\frac{2}{3}-\varphi \right)  - \frac{1+\xi}{3} \frac{{\cal R}^\prime_\Delta(\xi)}{{\cal R}(\xi)} \, .
\label{eq:tintiattractor}
\ee
The solution of this differential equation subject to the boundary condition $\varphi(0) = 3/4$ is shown in Fig.~\ref{fig:tinti_attractor}. As this figure shows, the moment method seems to reproduce the exact RTA attractor better than the Tinti matching principle.

\bibliography{attractor}

\begin{thebibliography}{85}%
\makeatletter
\providecommand \@ifxundefined [1]{%
 \@ifx{#1\undefined}
}%
\providecommand \@ifnum [1]{%
 \ifnum #1\expandafter \@firstoftwo
 \else \expandafter \@secondoftwo
 \fi
}%
\providecommand \@ifx [1]{%
 \ifx #1\expandafter \@firstoftwo
 \else \expandafter \@secondoftwo
 \fi
}%
\providecommand \natexlab [1]{#1}%
\providecommand \enquote  [1]{``#1''}%
\providecommand \bibnamefont  [1]{#1}%
\providecommand \bibfnamefont [1]{#1}%
\providecommand \citenamefont [1]{#1}%
\providecommand \href@noop [0]{\@secondoftwo}%
\providecommand \href [0]{\begingroup \@sanitize@url \@href}%
\providecommand \@href[1]{\@@startlink{#1}\@@href}%
\providecommand \@@href[1]{\endgroup#1\@@endlink}%
\providecommand \@sanitize@url [0]{\catcode `\\12\catcode `\$12\catcode
  `\&12\catcode `\#12\catcode `\^12\catcode `\_12\catcode `\%12\relax}%
\providecommand \@@startlink[1]{}%
\providecommand \@@endlink[0]{}%
\providecommand \url  [0]{\begingroup\@sanitize@url \@url }%
\providecommand \@url [1]{\endgroup\@href {#1}{\urlprefix }}%
\providecommand \urlprefix  [0]{URL }%
\providecommand \Eprint [0]{\href }%
\providecommand \doibase [0]{http://dx.doi.org/}%
\providecommand \selectlanguage [0]{\@gobble}%
\providecommand \bibinfo  [0]{\@secondoftwo}%
\providecommand \bibfield  [0]{\@secondoftwo}%
\providecommand \translation [1]{[#1]}%
\providecommand \BibitemOpen [0]{}%
\providecommand \bibitemStop [0]{}%
\providecommand \bibitemNoStop [0]{.\EOS\space}%
\providecommand \EOS [0]{\spacefactor3000\relax}%
\providecommand \BibitemShut  [1]{\csname bibitem#1\endcsname}%
\let\auto@bib@innerbib\@empty
\bibitem [{\citenamefont {Heinz}\ and\ \citenamefont
  {Snellings}(2013)}]{Heinz:2013th}%
  \BibitemOpen
  \bibfield  {author} {\bibinfo {author} {\bibfnamefont {U.}~\bibnamefont
  {Heinz}}\ and\ \bibinfo {author} {\bibfnamefont {R.}~\bibnamefont
  {Snellings}},\ }\href {\doibase 10.1146/annurev-nucl-102212-170540}
  {\bibfield  {journal} {\bibinfo  {journal} {Ann. Rev. Nucl. Part. Sci.}\
  }\textbf {\bibinfo {volume} {63}},\ \bibinfo {pages} {123} (\bibinfo {year}
  {2013})},\ \Eprint {http://arxiv.org/abs/1301.2826} {arXiv:1301.2826
  [nucl-th]} \BibitemShut {NoStop}%
\bibitem [{\citenamefont {Chapman}\ and\ \citenamefont
  {Cowling}(1952)}]{Chapman_Enskog}%
  \BibitemOpen
  \bibfield  {author} {\bibinfo {author} {\bibfnamefont {S.}~\bibnamefont
  {Chapman}}\ and\ \bibinfo {author} {\bibfnamefont {T.~G.}\ \bibnamefont
  {Cowling}},\ }\href@noop {} {\emph {\bibinfo {title} {The Mathematical Theory
  of Non-Uniform Gases}}}\ (\bibinfo  {publisher} {Cambridge University Press,
  Cambridge, England},\ \bibinfo {year} {1952})\BibitemShut {NoStop}%
\bibitem [{\citenamefont {Heller}\ \emph
  {et~al.}(2013{\natexlab{a}})\citenamefont {Heller}, \citenamefont {Janik},\
  and\ \citenamefont {Witaszczyk}}]{Heller:2013fn}%
  \BibitemOpen
  \bibfield  {author} {\bibinfo {author} {\bibfnamefont {M.~P.}\ \bibnamefont
  {Heller}}, \bibinfo {author} {\bibfnamefont {R.~A.}\ \bibnamefont {Janik}}, \
  and\ \bibinfo {author} {\bibfnamefont {P.}~\bibnamefont {Witaszczyk}},\
  }\href {\doibase 10.1103/PhysRevLett.110.211602} {\bibfield  {journal}
  {\bibinfo  {journal} {Phys. Rev. Lett.}\ }\textbf {\bibinfo {volume} {110}},\
  \bibinfo {pages} {211602} (\bibinfo {year} {2013}{\natexlab{a}})},\ \Eprint
  {http://arxiv.org/abs/1302.0697} {arXiv:1302.0697 [hep-th]} \BibitemShut
  {NoStop}%
\bibitem [{\citenamefont {Buchel}\ \emph {et~al.}(2016)\citenamefont {Buchel},
  \citenamefont {Heller},\ and\ \citenamefont {Noronha}}]{Buchel:2016cbj}%
  \BibitemOpen
  \bibfield  {author} {\bibinfo {author} {\bibfnamefont {A.}~\bibnamefont
  {Buchel}}, \bibinfo {author} {\bibfnamefont {M.~P.}\ \bibnamefont {Heller}},
  \ and\ \bibinfo {author} {\bibfnamefont {J.}~\bibnamefont {Noronha}},\ }\href
  {\doibase 10.1103/PhysRevD.94.106011} {\bibfield  {journal} {\bibinfo
  {journal} {Phys. Rev.}\ }\textbf {\bibinfo {volume} {D94}},\ \bibinfo {pages}
  {106011} (\bibinfo {year} {2016})},\ \Eprint
  {http://arxiv.org/abs/1603.05344} {arXiv:1603.05344 [hep-th]} \BibitemShut
  {NoStop}%
\bibitem [{\citenamefont {Heller}\ \emph {et~al.}(2016)\citenamefont {Heller},
  \citenamefont {Kurkela},\ and\ \citenamefont {Spalinski}}]{Heller:2016rtz}%
  \BibitemOpen
  \bibfield  {author} {\bibinfo {author} {\bibfnamefont {M.~P.}\ \bibnamefont
  {Heller}}, \bibinfo {author} {\bibfnamefont {A.}~\bibnamefont {Kurkela}}, \
  and\ \bibinfo {author} {\bibfnamefont {M.}~\bibnamefont {Spalinski}},\
  }\href@noop {} {\  (\bibinfo {year} {2016})},\ \Eprint
  {http://arxiv.org/abs/1609.04803} {arXiv:1609.04803 [nucl-th]} \BibitemShut
  {NoStop}%
\bibitem [{\citenamefont {Denicol}\ and\ \citenamefont
  {Noronha}(2016)}]{Denicol:2016bjh}%
  \BibitemOpen
  \bibfield  {author} {\bibinfo {author} {\bibfnamefont {G.~S.}\ \bibnamefont
  {Denicol}}\ and\ \bibinfo {author} {\bibfnamefont {J.}~\bibnamefont
  {Noronha}},\ }\href@noop {} {\  (\bibinfo {year} {2016})},\ \Eprint
  {http://arxiv.org/abs/1608.07869} {arXiv:1608.07869 [nucl-th]} \BibitemShut
  {NoStop}%
\bibitem [{\citenamefont {Heinz}(2004)}]{Heinz:2004qz}%
  \BibitemOpen
  \bibfield  {author} {\bibinfo {author} {\bibfnamefont {U.~W.}\ \bibnamefont
  {Heinz}},\ }in\ \href {http://doc.cern.ch/yellowrep/CERN-2004-001} {\emph
  {\bibinfo {booktitle} {{2002 European School of high-energy physics, Pylos,
  Greece, 25 Aug-7 Sep 2002: Proceedings}}}}\ (\bibinfo {year} {2004})\ pp.\
  \bibinfo {pages} {165--238},\ \Eprint {http://arxiv.org/abs/hep-ph/0407360}
  {arXiv:hep-ph/0407360 [hep-ph]} \BibitemShut {NoStop}%
\bibitem [{\citenamefont {Chesler}\ and\ \citenamefont
  {Yaffe}(2009)}]{Chesler:2008hg}%
  \BibitemOpen
  \bibfield  {author} {\bibinfo {author} {\bibfnamefont {P.~M.}\ \bibnamefont
  {Chesler}}\ and\ \bibinfo {author} {\bibfnamefont {L.~G.}\ \bibnamefont
  {Yaffe}},\ }\href {\doibase 10.1103/PhysRevLett.102.211601} {\bibfield
  {journal} {\bibinfo  {journal} {Phys. Rev. Lett.}\ }\textbf {\bibinfo
  {volume} {102}},\ \bibinfo {pages} {211601} (\bibinfo {year} {2009})},\
  \Eprint {http://arxiv.org/abs/0812.2053} {arXiv:0812.2053 [hep-th]}
  \BibitemShut {NoStop}%
\bibitem [{\citenamefont {Beuf}\ \emph {et~al.}(2009)\citenamefont {Beuf},
  \citenamefont {Heller}, \citenamefont {Janik},\ and\ \citenamefont
  {Peschanski}}]{Beuf:2009cx}%
  \BibitemOpen
  \bibfield  {author} {\bibinfo {author} {\bibfnamefont {G.}~\bibnamefont
  {Beuf}}, \bibinfo {author} {\bibfnamefont {M.~P.}\ \bibnamefont {Heller}},
  \bibinfo {author} {\bibfnamefont {R.~A.}\ \bibnamefont {Janik}}, \ and\
  \bibinfo {author} {\bibfnamefont {R.}~\bibnamefont {Peschanski}},\ }\href
  {\doibase 10.1088/1126-6708/2009/10/043} {\bibfield  {journal} {\bibinfo
  {journal} {JHEP}\ }\textbf {\bibinfo {volume} {10}},\ \bibinfo {pages} {043}
  (\bibinfo {year} {2009})},\ \Eprint {http://arxiv.org/abs/0906.4423}
  {arXiv:0906.4423 [hep-th]} \BibitemShut {NoStop}%
\bibitem [{\citenamefont {Chesler}\ and\ \citenamefont
  {Yaffe}(2010)}]{Chesler:2009cy}%
  \BibitemOpen
  \bibfield  {author} {\bibinfo {author} {\bibfnamefont {P.~M.}\ \bibnamefont
  {Chesler}}\ and\ \bibinfo {author} {\bibfnamefont {L.~G.}\ \bibnamefont
  {Yaffe}},\ }\href {\doibase 10.1103/PhysRevD.82.026006} {\bibfield  {journal}
  {\bibinfo  {journal} {Phys. Rev.}\ }\textbf {\bibinfo {volume} {D82}},\
  \bibinfo {pages} {026006} (\bibinfo {year} {2010})},\ \Eprint
  {http://arxiv.org/abs/0906.4426} {arXiv:0906.4426 [hep-th]} \BibitemShut
  {NoStop}%
\bibitem [{\citenamefont {Heller}\ \emph
  {et~al.}(2012{\natexlab{a}})\citenamefont {Heller}, \citenamefont {Janik},\
  and\ \citenamefont {Witaszczyk}}]{Heller:2011ju}%
  \BibitemOpen
  \bibfield  {author} {\bibinfo {author} {\bibfnamefont {M.~P.}\ \bibnamefont
  {Heller}}, \bibinfo {author} {\bibfnamefont {R.~A.}\ \bibnamefont {Janik}}, \
  and\ \bibinfo {author} {\bibfnamefont {P.}~\bibnamefont {Witaszczyk}},\
  }\href {\doibase 10.1103/PhysRevLett.108.201602} {\bibfield  {journal}
  {\bibinfo  {journal} {Phys. Rev. Lett.}\ }\textbf {\bibinfo {volume} {108}},\
  \bibinfo {pages} {201602} (\bibinfo {year} {2012}{\natexlab{a}})},\ \Eprint
  {http://arxiv.org/abs/1103.3452} {arXiv:1103.3452 [hep-th]} \BibitemShut
  {NoStop}%
\bibitem [{\citenamefont {Heller}\ \emph
  {et~al.}(2012{\natexlab{b}})\citenamefont {Heller}, \citenamefont {Janik},\
  and\ \citenamefont {Witaszczyk}}]{Heller:2012je}%
  \BibitemOpen
  \bibfield  {author} {\bibinfo {author} {\bibfnamefont {M.~P.}\ \bibnamefont
  {Heller}}, \bibinfo {author} {\bibfnamefont {R.~A.}\ \bibnamefont {Janik}}, \
  and\ \bibinfo {author} {\bibfnamefont {P.}~\bibnamefont {Witaszczyk}},\
  }\href {\doibase 10.1103/PhysRevD.85.126002} {\bibfield  {journal} {\bibinfo
  {journal} {Phys. Rev.}\ }\textbf {\bibinfo {volume} {D85}},\ \bibinfo {pages}
  {126002} (\bibinfo {year} {2012}{\natexlab{b}})},\ \Eprint
  {http://arxiv.org/abs/1203.0755} {arXiv:1203.0755 [hep-th]} \BibitemShut
  {NoStop}%
\bibitem [{\citenamefont {Heller}\ \emph
  {et~al.}(2012{\natexlab{c}})\citenamefont {Heller}, \citenamefont {Mateos},
  \citenamefont {van~der Schee},\ and\ \citenamefont
  {Trancanelli}}]{Heller:2012km}%
  \BibitemOpen
  \bibfield  {author} {\bibinfo {author} {\bibfnamefont {M.~P.}\ \bibnamefont
  {Heller}}, \bibinfo {author} {\bibfnamefont {D.}~\bibnamefont {Mateos}},
  \bibinfo {author} {\bibfnamefont {W.}~\bibnamefont {van~der Schee}}, \ and\
  \bibinfo {author} {\bibfnamefont {D.}~\bibnamefont {Trancanelli}},\ }\href
  {\doibase 10.1103/PhysRevLett.108.191601} {\bibfield  {journal} {\bibinfo
  {journal} {Phys. Rev. Lett.}\ }\textbf {\bibinfo {volume} {108}},\ \bibinfo
  {pages} {191601} (\bibinfo {year} {2012}{\natexlab{c}})},\ \Eprint
  {http://arxiv.org/abs/1202.0981} {arXiv:1202.0981 [hep-th]} \BibitemShut
  {NoStop}%
\bibitem [{\citenamefont {van~der Schee}(2013)}]{vanderSchee:2012qj}%
  \BibitemOpen
  \bibfield  {author} {\bibinfo {author} {\bibfnamefont {W.}~\bibnamefont
  {van~der Schee}},\ }\href {\doibase 10.1103/PhysRevD.87.061901} {\bibfield
  {journal} {\bibinfo  {journal} {Phys. Rev.}\ }\textbf {\bibinfo {volume}
  {D87}},\ \bibinfo {pages} {061901} (\bibinfo {year} {2013})},\ \Eprint
  {http://arxiv.org/abs/1211.2218} {arXiv:1211.2218 [hep-th]} \BibitemShut
  {NoStop}%
\bibitem [{\citenamefont {Casalderrey-Solana}\ \emph
  {et~al.}(2013)\citenamefont {Casalderrey-Solana}, \citenamefont {Heller},
  \citenamefont {Mateos},\ and\ \citenamefont {van~der
  Schee}}]{Casalderrey-Solana:2013aba}%
  \BibitemOpen
  \bibfield  {author} {\bibinfo {author} {\bibfnamefont {J.}~\bibnamefont
  {Casalderrey-Solana}}, \bibinfo {author} {\bibfnamefont {M.~P.}\ \bibnamefont
  {Heller}}, \bibinfo {author} {\bibfnamefont {D.}~\bibnamefont {Mateos}}, \
  and\ \bibinfo {author} {\bibfnamefont {W.}~\bibnamefont {van~der Schee}},\
  }\href {\doibase 10.1103/PhysRevLett.111.181601} {\bibfield  {journal}
  {\bibinfo  {journal} {Phys. Rev. Lett.}\ }\textbf {\bibinfo {volume} {111}},\
  \bibinfo {pages} {181601} (\bibinfo {year} {2013})},\ \Eprint
  {http://arxiv.org/abs/1305.4919} {arXiv:1305.4919 [hep-th]} \BibitemShut
  {NoStop}%
\bibitem [{\citenamefont {van~der Schee}\ \emph {et~al.}(2013)\citenamefont
  {van~der Schee}, \citenamefont {Romatschke},\ and\ \citenamefont
  {Pratt}}]{vanderSchee:2013pia}%
  \BibitemOpen
  \bibfield  {author} {\bibinfo {author} {\bibfnamefont {W.}~\bibnamefont
  {van~der Schee}}, \bibinfo {author} {\bibfnamefont {P.}~\bibnamefont
  {Romatschke}}, \ and\ \bibinfo {author} {\bibfnamefont {S.}~\bibnamefont
  {Pratt}},\ }\href {\doibase 10.1103/PhysRevLett.111.222302} {\bibfield
  {journal} {\bibinfo  {journal} {Phys. Rev. Lett.}\ }\textbf {\bibinfo
  {volume} {111}},\ \bibinfo {pages} {222302} (\bibinfo {year} {2013})},\
  \Eprint {http://arxiv.org/abs/1307.2539} {arXiv:1307.2539 [nucl-th]}
  \BibitemShut {NoStop}%
\bibitem [{\citenamefont {Heller}\ \emph
  {et~al.}(2013{\natexlab{b}})\citenamefont {Heller}, \citenamefont {Mateos},
  \citenamefont {van~der Schee},\ and\ \citenamefont
  {Triana}}]{Heller:2013oxa}%
  \BibitemOpen
  \bibfield  {author} {\bibinfo {author} {\bibfnamefont {M.~P.}\ \bibnamefont
  {Heller}}, \bibinfo {author} {\bibfnamefont {D.}~\bibnamefont {Mateos}},
  \bibinfo {author} {\bibfnamefont {W.}~\bibnamefont {van~der Schee}}, \ and\
  \bibinfo {author} {\bibfnamefont {M.}~\bibnamefont {Triana}},\ }\href
  {\doibase 10.1007/JHEP09(2013)026} {\bibfield  {journal} {\bibinfo  {journal}
  {JHEP}\ }\textbf {\bibinfo {volume} {09}},\ \bibinfo {pages} {026} (\bibinfo
  {year} {2013}{\natexlab{b}})},\ \Eprint {http://arxiv.org/abs/1304.5172}
  {arXiv:1304.5172 [hep-th]} \BibitemShut {NoStop}%
\bibitem [{\citenamefont {Keegan}\ \emph {et~al.}(2016)\citenamefont {Keegan},
  \citenamefont {Kurkela}, \citenamefont {Romatschke}, \citenamefont {van~der
  Schee},\ and\ \citenamefont {Zhu}}]{Keegan:2015avk}%
  \BibitemOpen
  \bibfield  {author} {\bibinfo {author} {\bibfnamefont {L.}~\bibnamefont
  {Keegan}}, \bibinfo {author} {\bibfnamefont {A.}~\bibnamefont {Kurkela}},
  \bibinfo {author} {\bibfnamefont {P.}~\bibnamefont {Romatschke}}, \bibinfo
  {author} {\bibfnamefont {W.}~\bibnamefont {van~der Schee}}, \ and\ \bibinfo
  {author} {\bibfnamefont {Y.}~\bibnamefont {Zhu}},\ }\href {\doibase
  10.1007/JHEP04(2016)031} {\bibfield  {journal} {\bibinfo  {journal} {JHEP}\
  }\textbf {\bibinfo {volume} {04}},\ \bibinfo {pages} {031} (\bibinfo {year}
  {2016})},\ \Eprint {http://arxiv.org/abs/1512.05347} {arXiv:1512.05347
  [hep-th]} \BibitemShut {NoStop}%
\bibitem [{\citenamefont {Chesler}(2015)}]{Chesler:2015bba}%
  \BibitemOpen
  \bibfield  {author} {\bibinfo {author} {\bibfnamefont {P.~M.}\ \bibnamefont
  {Chesler}},\ }\href {\doibase 10.1103/PhysRevLett.115.241602} {\bibfield
  {journal} {\bibinfo  {journal} {Phys. Rev. Lett.}\ }\textbf {\bibinfo
  {volume} {115}},\ \bibinfo {pages} {241602} (\bibinfo {year} {2015})},\
  \Eprint {http://arxiv.org/abs/1506.02209} {arXiv:1506.02209 [hep-th]}
  \BibitemShut {NoStop}%
\bibitem [{\citenamefont {Kurkela}\ and\ \citenamefont
  {Zhu}(2015)}]{Kurkela:2015qoa}%
  \BibitemOpen
  \bibfield  {author} {\bibinfo {author} {\bibfnamefont {A.}~\bibnamefont
  {Kurkela}}\ and\ \bibinfo {author} {\bibfnamefont {Y.}~\bibnamefont {Zhu}},\
  }\href {\doibase 10.1103/PhysRevLett.115.182301} {\bibfield  {journal}
  {\bibinfo  {journal} {Phys. Rev. Lett.}\ }\textbf {\bibinfo {volume} {115}},\
  \bibinfo {pages} {182301} (\bibinfo {year} {2015})},\ \Eprint
  {http://arxiv.org/abs/1506.06647} {arXiv:1506.06647 [hep-ph]} \BibitemShut
  {NoStop}%
\bibitem [{\citenamefont {Chesler}(2016)}]{Chesler:2016ceu}%
  \BibitemOpen
  \bibfield  {author} {\bibinfo {author} {\bibfnamefont {P.~M.}\ \bibnamefont
  {Chesler}},\ }\href {\doibase 10.1007/JHEP03(2016)146} {\bibfield  {journal}
  {\bibinfo  {journal} {JHEP}\ }\textbf {\bibinfo {volume} {03}},\ \bibinfo
  {pages} {146} (\bibinfo {year} {2016})},\ \Eprint
  {http://arxiv.org/abs/1601.01583} {arXiv:1601.01583 [hep-th]} \BibitemShut
  {NoStop}%
\bibitem [{\citenamefont {Attems}\ \emph {et~al.}(2016)\citenamefont {Attems},
  \citenamefont {Casalderrey-Solana}, \citenamefont {Mateos}, \citenamefont
  {Papadimitriou}, \citenamefont {Santos-Oliván}, \citenamefont {Sopuerta},
  \citenamefont {Triana},\ and\ \citenamefont {Zilhão}}]{Attems:2016ugt}%
  \BibitemOpen
  \bibfield  {author} {\bibinfo {author} {\bibfnamefont {M.}~\bibnamefont
  {Attems}}, \bibinfo {author} {\bibfnamefont {J.}~\bibnamefont
  {Casalderrey-Solana}}, \bibinfo {author} {\bibfnamefont {D.}~\bibnamefont
  {Mateos}}, \bibinfo {author} {\bibfnamefont {I.}~\bibnamefont
  {Papadimitriou}}, \bibinfo {author} {\bibfnamefont {D.}~\bibnamefont
  {Santos-Oliván}}, \bibinfo {author} {\bibfnamefont {C.~F.}\ \bibnamefont
  {Sopuerta}}, \bibinfo {author} {\bibfnamefont {M.}~\bibnamefont {Triana}}, \
  and\ \bibinfo {author} {\bibfnamefont {M.}~\bibnamefont {Zilhão}},\ }\href
  {\doibase 10.1007/JHEP10(2016)155} {\bibfield  {journal} {\bibinfo  {journal}
  {JHEP}\ }\textbf {\bibinfo {volume} {10}},\ \bibinfo {pages} {155} (\bibinfo
  {year} {2016})},\ \Eprint {http://arxiv.org/abs/1603.01254} {arXiv:1603.01254
  [hep-th]} \BibitemShut {NoStop}%
\bibitem [{\citenamefont {Attems}\ \emph
  {et~al.}(2017{\natexlab{a}})\citenamefont {Attems}, \citenamefont
  {Casalderrey-Solana}, \citenamefont {Mateos}, \citenamefont {Santos-Oliván},
  \citenamefont {Sopuerta}, \citenamefont {Triana},\ and\ \citenamefont
  {Zilhão}}]{Attems:2016tby}%
  \BibitemOpen
  \bibfield  {author} {\bibinfo {author} {\bibfnamefont {M.}~\bibnamefont
  {Attems}}, \bibinfo {author} {\bibfnamefont {J.}~\bibnamefont
  {Casalderrey-Solana}}, \bibinfo {author} {\bibfnamefont {D.}~\bibnamefont
  {Mateos}}, \bibinfo {author} {\bibfnamefont {D.}~\bibnamefont
  {Santos-Oliván}}, \bibinfo {author} {\bibfnamefont {C.~F.}\ \bibnamefont
  {Sopuerta}}, \bibinfo {author} {\bibfnamefont {M.}~\bibnamefont {Triana}}, \
  and\ \bibinfo {author} {\bibfnamefont {M.}~\bibnamefont {Zilhão}},\ }\href
  {\doibase 10.1007/JHEP01(2017)026} {\bibfield  {journal} {\bibinfo  {journal}
  {JHEP}\ }\textbf {\bibinfo {volume} {01}},\ \bibinfo {pages} {026} (\bibinfo
  {year} {2017}{\natexlab{a}})},\ \Eprint {http://arxiv.org/abs/1604.06439}
  {arXiv:1604.06439 [hep-th]} \BibitemShut {NoStop}%
\bibitem [{\citenamefont {Attems}\ \emph
  {et~al.}(2017{\natexlab{b}})\citenamefont {Attems}, \citenamefont
  {Casalderrey-Solana}, \citenamefont {Mateos}, \citenamefont {Santos-Oliván},
  \citenamefont {Sopuerta}, \citenamefont {Triana},\ and\ \citenamefont
  {Zilhão}}]{Attems:2017zam}%
  \BibitemOpen
  \bibfield  {author} {\bibinfo {author} {\bibfnamefont {M.}~\bibnamefont
  {Attems}}, \bibinfo {author} {\bibfnamefont {J.}~\bibnamefont
  {Casalderrey-Solana}}, \bibinfo {author} {\bibfnamefont {D.}~\bibnamefont
  {Mateos}}, \bibinfo {author} {\bibfnamefont {D.}~\bibnamefont
  {Santos-Oliván}}, \bibinfo {author} {\bibfnamefont {C.~F.}\ \bibnamefont
  {Sopuerta}}, \bibinfo {author} {\bibfnamefont {M.}~\bibnamefont {Triana}}, \
  and\ \bibinfo {author} {\bibfnamefont {M.}~\bibnamefont {Zilhão}},\ }\href
  {\doibase 10.1007/JHEP06(2017)154} {\bibfield  {journal} {\bibinfo  {journal}
  {JHEP}\ }\textbf {\bibinfo {volume} {06}},\ \bibinfo {pages} {154} (\bibinfo
  {year} {2017}{\natexlab{b}})},\ \Eprint {http://arxiv.org/abs/1703.09681}
  {arXiv:1703.09681 [hep-th]} \BibitemShut {NoStop}%
\bibitem [{\citenamefont {Florkowski}\ \emph {et~al.}(2017)\citenamefont
  {Florkowski}, \citenamefont {Heller},\ and\ \citenamefont
  {Spalinski}}]{Florkowski:2017olj}%
  \BibitemOpen
  \bibfield  {author} {\bibinfo {author} {\bibfnamefont {W.}~\bibnamefont
  {Florkowski}}, \bibinfo {author} {\bibfnamefont {M.~P.}\ \bibnamefont
  {Heller}}, \ and\ \bibinfo {author} {\bibfnamefont {M.}~\bibnamefont
  {Spalinski}},\ }\href@noop {} {\  (\bibinfo {year} {2017})},\ \Eprint
  {http://arxiv.org/abs/1707.02282} {arXiv:1707.02282 [hep-ph]} \BibitemShut
  {NoStop}%
\bibitem [{\citenamefont {Florkowski}\ and\ \citenamefont
  {Ryblewski}(2011)}]{Florkowski:2010cf}%
  \BibitemOpen
  \bibfield  {author} {\bibinfo {author} {\bibfnamefont {W.}~\bibnamefont
  {Florkowski}}\ and\ \bibinfo {author} {\bibfnamefont {R.}~\bibnamefont
  {Ryblewski}},\ }\href {\doibase 10.1103/PhysRevC.83.034907} {\bibfield
  {journal} {\bibinfo  {journal} {Phys.Rev.}\ }\textbf {\bibinfo {volume}
  {C83}},\ \bibinfo {pages} {034907} (\bibinfo {year} {2011})},\ \Eprint
  {http://arxiv.org/abs/1007.0130} {arXiv:1007.0130 [nucl-th]} \BibitemShut
  {NoStop}%
\bibitem [{\citenamefont {Martinez}\ and\ \citenamefont
  {Strickland}(2010)}]{Martinez:2010sc}%
  \BibitemOpen
  \bibfield  {author} {\bibinfo {author} {\bibfnamefont {M.}~\bibnamefont
  {Martinez}}\ and\ \bibinfo {author} {\bibfnamefont {M.}~\bibnamefont
  {Strickland}},\ }\href {\doibase 10.1016/j.nuclphysa.2010.08.011} {\bibfield
  {journal} {\bibinfo  {journal} {Nucl. Phys.}\ }\textbf {\bibinfo {volume}
  {A848}},\ \bibinfo {pages} {183} (\bibinfo {year} {2010})},\ \Eprint
  {http://arxiv.org/abs/1007.0889} {arXiv:1007.0889 [nucl-th]} \BibitemShut
  {NoStop}%
\bibitem [{\citenamefont {Ryblewski}\ and\ \citenamefont
  {Florkowski}(2011)}]{Ryblewski:2010ch}%
  \BibitemOpen
  \bibfield  {author} {\bibinfo {author} {\bibfnamefont {R.}~\bibnamefont
  {Ryblewski}}\ and\ \bibinfo {author} {\bibfnamefont {W.}~\bibnamefont
  {Florkowski}},\ }\href {\doibase 10.5506/APhysPolB.42.115} {\bibfield
  {journal} {\bibinfo  {journal} {Acta Phys. Polon.}\ }\textbf {\bibinfo
  {volume} {B42}},\ \bibinfo {pages} {115} (\bibinfo {year} {2011})},\ \Eprint
  {http://arxiv.org/abs/1011.6213} {arXiv:1011.6213 [nucl-th]} \BibitemShut
  {NoStop}%
\bibitem [{\citenamefont {Martinez}\ \emph {et~al.}(2012)\citenamefont
  {Martinez}, \citenamefont {Ryblewski},\ and\ \citenamefont
  {Strickland}}]{Martinez:2012tu}%
  \BibitemOpen
  \bibfield  {author} {\bibinfo {author} {\bibfnamefont {M.}~\bibnamefont
  {Martinez}}, \bibinfo {author} {\bibfnamefont {R.}~\bibnamefont {Ryblewski}},
  \ and\ \bibinfo {author} {\bibfnamefont {M.}~\bibnamefont {Strickland}},\
  }\href@noop {} {\bibfield  {journal} {\bibinfo  {journal} {Phys.Rev.}\
  }\textbf {\bibinfo {volume} {C85}},\ \bibinfo {pages} {064913} (\bibinfo
  {year} {2012})},\ \Eprint {http://arxiv.org/abs/1204.1473} {arXiv:1204.1473
  [nucl-th]} \BibitemShut {NoStop}%
\bibitem [{\citenamefont {Ryblewski}\ and\ \citenamefont
  {Florkowski}(2012)}]{Ryblewski:2012rr}%
  \BibitemOpen
  \bibfield  {author} {\bibinfo {author} {\bibfnamefont {R.}~\bibnamefont
  {Ryblewski}}\ and\ \bibinfo {author} {\bibfnamefont {W.}~\bibnamefont
  {Florkowski}},\ }\href {\doibase 10.1103/PhysRevC.85.064901} {\bibfield
  {journal} {\bibinfo  {journal} {Phys. Rev.}\ }\textbf {\bibinfo {volume}
  {C85}},\ \bibinfo {pages} {064901} (\bibinfo {year} {2012})},\ \Eprint
  {http://arxiv.org/abs/1204.2624} {arXiv:1204.2624 [nucl-th]} \BibitemShut
  {NoStop}%
\bibitem [{\citenamefont {Bazow}\ \emph {et~al.}(2014)\citenamefont {Bazow},
  \citenamefont {Heinz},\ and\ \citenamefont {Strickland}}]{Bazow:2013ifa}%
  \BibitemOpen
  \bibfield  {author} {\bibinfo {author} {\bibfnamefont {D.}~\bibnamefont
  {Bazow}}, \bibinfo {author} {\bibfnamefont {U.~W.}\ \bibnamefont {Heinz}}, \
  and\ \bibinfo {author} {\bibfnamefont {M.}~\bibnamefont {Strickland}},\
  }\href {\doibase 10.1103/PhysRevC.90.054910} {\bibfield  {journal} {\bibinfo
  {journal} {Phys.Rev.}\ }\textbf {\bibinfo {volume} {C90}},\ \bibinfo {pages}
  {054910} (\bibinfo {year} {2014})},\ \Eprint {http://arxiv.org/abs/1311.6720}
  {arXiv:1311.6720 [nucl-th]} \BibitemShut {NoStop}%
\bibitem [{\citenamefont {Tinti}\ and\ \citenamefont
  {Florkowski}(2014)}]{Tinti:2013vba}%
  \BibitemOpen
  \bibfield  {author} {\bibinfo {author} {\bibfnamefont {L.}~\bibnamefont
  {Tinti}}\ and\ \bibinfo {author} {\bibfnamefont {W.}~\bibnamefont
  {Florkowski}},\ }\href {\doibase 10.1103/PhysRevC.89.034907} {\bibfield
  {journal} {\bibinfo  {journal} {Phys.Rev.}\ }\textbf {\bibinfo {volume}
  {C89}},\ \bibinfo {pages} {034907} (\bibinfo {year} {2014})},\ \Eprint
  {http://arxiv.org/abs/1312.6614} {arXiv:1312.6614 [nucl-th]} \BibitemShut
  {NoStop}%
\bibitem [{\citenamefont {Nopoush}\ \emph {et~al.}(2014)\citenamefont
  {Nopoush}, \citenamefont {Ryblewski},\ and\ \citenamefont
  {Strickland}}]{Nopoush:2014pfa}%
  \BibitemOpen
  \bibfield  {author} {\bibinfo {author} {\bibfnamefont {M.}~\bibnamefont
  {Nopoush}}, \bibinfo {author} {\bibfnamefont {R.}~\bibnamefont {Ryblewski}},
  \ and\ \bibinfo {author} {\bibfnamefont {M.}~\bibnamefont {Strickland}},\
  }\href {\doibase 10.1103/PhysRevC.90.014908} {\bibfield  {journal} {\bibinfo
  {journal} {Phys.Rev.}\ }\textbf {\bibinfo {volume} {C90}},\ \bibinfo {pages}
  {014908} (\bibinfo {year} {2014})},\ \Eprint {http://arxiv.org/abs/1405.1355}
  {arXiv:1405.1355 [hep-ph]} \BibitemShut {NoStop}%
\bibitem [{\citenamefont {Tinti}(2016)}]{Tinti:2015xwa}%
  \BibitemOpen
  \bibfield  {author} {\bibinfo {author} {\bibfnamefont {L.}~\bibnamefont
  {Tinti}},\ }\href {\doibase 10.1103/PhysRevC.94.044902} {\bibfield  {journal}
  {\bibinfo  {journal} {Phys. Rev.}\ }\textbf {\bibinfo {volume} {C94}},\
  \bibinfo {pages} {044902} (\bibinfo {year} {2016})},\ \Eprint
  {http://arxiv.org/abs/1506.07164} {arXiv:1506.07164 [hep-ph]} \BibitemShut
  {NoStop}%
\bibitem [{\citenamefont {Bazow}\ \emph
  {et~al.}(2015{\natexlab{a}})\citenamefont {Bazow}, \citenamefont {Heinz},\
  and\ \citenamefont {Martinez}}]{Bazow:2015cha}%
  \BibitemOpen
  \bibfield  {author} {\bibinfo {author} {\bibfnamefont {D.}~\bibnamefont
  {Bazow}}, \bibinfo {author} {\bibfnamefont {U.~W.}\ \bibnamefont {Heinz}}, \
  and\ \bibinfo {author} {\bibfnamefont {M.}~\bibnamefont {Martinez}},\ }\href
  {\doibase http://dx.doi.org/10.1103/PhysRevC.91.064903} {\bibfield  {journal}
  {\bibinfo  {journal} {Phys.Rev.}\ }\textbf {\bibinfo {volume} {C91}},\
  \bibinfo {pages} {064903} (\bibinfo {year} {2015}{\natexlab{a}})},\ \Eprint
  {http://arxiv.org/abs/1503.07443} {arXiv:1503.07443 [nucl-th]} \BibitemShut
  {NoStop}%
\bibitem [{\citenamefont {Strickland}\ \emph {et~al.}(2016)\citenamefont
  {Strickland}, \citenamefont {Nopoush},\ and\ \citenamefont
  {Ryblewski}}]{Strickland:2015utc}%
  \BibitemOpen
  \bibfield  {author} {\bibinfo {author} {\bibfnamefont {M.}~\bibnamefont
  {Strickland}}, \bibinfo {author} {\bibfnamefont {M.}~\bibnamefont {Nopoush}},
  \ and\ \bibinfo {author} {\bibfnamefont {R.}~\bibnamefont {Ryblewski}},\
  }\bibfield  {booktitle} {\emph {\bibinfo {booktitle} {{Proceedings, 25th
  International Conference on Ultra-Relativistic Nucleus-Nucleus Collisions
  (Quark Matter 2015): Kobe, Japan, September 27-October 3, 2015}}},\ }\href
  {\doibase 10.1016/j.nuclphysa.2016.02.014} {\bibfield  {journal} {\bibinfo
  {journal} {Nucl. Phys.}\ }\textbf {\bibinfo {volume} {A956}},\ \bibinfo
  {pages} {268} (\bibinfo {year} {2016})},\ \Eprint
  {http://arxiv.org/abs/1512.07334} {arXiv:1512.07334 [nucl-th]} \BibitemShut
  {NoStop}%
\bibitem [{\citenamefont {Alqahtani}\ \emph {et~al.}(2015)\citenamefont
  {Alqahtani}, \citenamefont {Nopoush},\ and\ \citenamefont
  {Strickland}}]{Alqahtani:2015qja}%
  \BibitemOpen
  \bibfield  {author} {\bibinfo {author} {\bibfnamefont {M.}~\bibnamefont
  {Alqahtani}}, \bibinfo {author} {\bibfnamefont {M.}~\bibnamefont {Nopoush}},
  \ and\ \bibinfo {author} {\bibfnamefont {M.}~\bibnamefont {Strickland}},\
  }\href {\doibase 10.1103/PhysRevC.92.054910} {\bibfield  {journal} {\bibinfo
  {journal} {Phys. Rev.}\ }\textbf {\bibinfo {volume} {C92}},\ \bibinfo {pages}
  {054910} (\bibinfo {year} {2015})},\ \Eprint
  {http://arxiv.org/abs/1509.02913} {arXiv:1509.02913 [hep-ph]} \BibitemShut
  {NoStop}%
\bibitem [{\citenamefont {Molnar}\ \emph
  {et~al.}(2016{\natexlab{a}})\citenamefont {Molnar}, \citenamefont {Niemi},\
  and\ \citenamefont {Rischke}}]{Molnar:2016vvu}%
  \BibitemOpen
  \bibfield  {author} {\bibinfo {author} {\bibfnamefont {E.}~\bibnamefont
  {Molnar}}, \bibinfo {author} {\bibfnamefont {H.}~\bibnamefont {Niemi}}, \
  and\ \bibinfo {author} {\bibfnamefont {D.~H.}\ \bibnamefont {Rischke}},\
  }\href {\doibase 10.1103/PhysRevD.93.114025} {\bibfield  {journal} {\bibinfo
  {journal} {Phys. Rev.}\ }\textbf {\bibinfo {volume} {D93}},\ \bibinfo {pages}
  {114025} (\bibinfo {year} {2016}{\natexlab{a}})},\ \Eprint
  {http://arxiv.org/abs/1602.00573} {arXiv:1602.00573 [nucl-th]} \BibitemShut
  {NoStop}%
\bibitem [{\citenamefont {Molnar}\ \emph
  {et~al.}(2016{\natexlab{b}})\citenamefont {Molnar}, \citenamefont {Niemi},\
  and\ \citenamefont {Rischke}}]{Molnar:2016gwq}%
  \BibitemOpen
  \bibfield  {author} {\bibinfo {author} {\bibfnamefont {E.}~\bibnamefont
  {Molnar}}, \bibinfo {author} {\bibfnamefont {H.}~\bibnamefont {Niemi}}, \
  and\ \bibinfo {author} {\bibfnamefont {D.~H.}\ \bibnamefont {Rischke}},\
  }\href {\doibase 10.1103/PhysRevD.94.125003} {\bibfield  {journal} {\bibinfo
  {journal} {Phys. Rev.}\ }\textbf {\bibinfo {volume} {D94}},\ \bibinfo {pages}
  {125003} (\bibinfo {year} {2016}{\natexlab{b}})},\ \Eprint
  {http://arxiv.org/abs/1606.09019} {arXiv:1606.09019 [nucl-th]} \BibitemShut
  {NoStop}%
\bibitem [{\citenamefont {Alqahtani}\ \emph
  {et~al.}(2017{\natexlab{a}})\citenamefont {Alqahtani}, \citenamefont
  {Nopoush},\ and\ \citenamefont {Strickland}}]{Alqahtani:2016rth}%
  \BibitemOpen
  \bibfield  {author} {\bibinfo {author} {\bibfnamefont {M.}~\bibnamefont
  {Alqahtani}}, \bibinfo {author} {\bibfnamefont {M.}~\bibnamefont {Nopoush}},
  \ and\ \bibinfo {author} {\bibfnamefont {M.}~\bibnamefont {Strickland}},\
  }\href {\doibase 10.1103/PhysRevC.95.034906} {\bibfield  {journal} {\bibinfo
  {journal} {Phys. Rev.}\ }\textbf {\bibinfo {volume} {C95}},\ \bibinfo {pages}
  {034906} (\bibinfo {year} {2017}{\natexlab{a}})},\ \Eprint
  {http://arxiv.org/abs/1605.02101} {arXiv:1605.02101 [nucl-th]} \BibitemShut
  {NoStop}%
\bibitem [{\citenamefont {Bluhm}\ and\ \citenamefont
  {Schaefer}(2015)}]{Bluhm:2015raa}%
  \BibitemOpen
  \bibfield  {author} {\bibinfo {author} {\bibfnamefont {M.}~\bibnamefont
  {Bluhm}}\ and\ \bibinfo {author} {\bibfnamefont {T.}~\bibnamefont
  {Schaefer}},\ }\href {\doibase 10.1103/PhysRevA.92.043602} {\bibfield
  {journal} {\bibinfo  {journal} {Phys. Rev.}\ }\textbf {\bibinfo {volume}
  {A92}},\ \bibinfo {pages} {043602} (\bibinfo {year} {2015})},\ \Eprint
  {http://arxiv.org/abs/1505.00846} {arXiv:1505.00846 [cond-mat.quant-gas]}
  \BibitemShut {NoStop}%
\bibitem [{\citenamefont {Bluhm}\ and\ \citenamefont
  {Schaefer}(2016)}]{Bluhm:2015bzi}%
  \BibitemOpen
  \bibfield  {author} {\bibinfo {author} {\bibfnamefont {M.}~\bibnamefont
  {Bluhm}}\ and\ \bibinfo {author} {\bibfnamefont {T.}~\bibnamefont
  {Schaefer}},\ }\href {\doibase 10.1103/PhysRevLett.116.115301} {\bibfield
  {journal} {\bibinfo  {journal} {Phys. Rev. Lett.}\ }\textbf {\bibinfo
  {volume} {116}},\ \bibinfo {pages} {115301} (\bibinfo {year} {2016})},\
  \Eprint {http://arxiv.org/abs/1512.00862} {arXiv:1512.00862
  [cond-mat.quant-gas]} \BibitemShut {NoStop}%
\bibitem [{\citenamefont {Alqahtani}\ \emph
  {et~al.}(2017{\natexlab{b}})\citenamefont {Alqahtani}, \citenamefont
  {Nopoush}, \citenamefont {Ryblewski},\ and\ \citenamefont
  {Strickland}}]{Alqahtani:2017jwl}%
  \BibitemOpen
  \bibfield  {author} {\bibinfo {author} {\bibfnamefont {M.}~\bibnamefont
  {Alqahtani}}, \bibinfo {author} {\bibfnamefont {M.}~\bibnamefont {Nopoush}},
  \bibinfo {author} {\bibfnamefont {R.}~\bibnamefont {Ryblewski}}, \ and\
  \bibinfo {author} {\bibfnamefont {M.}~\bibnamefont {Strickland}},\ }\href
  {\doibase 10.1103/PhysRevLett.119.042301} {\bibfield  {journal} {\bibinfo
  {journal} {Phys. Rev. Lett.}\ }\textbf {\bibinfo {volume} {119}},\ \bibinfo
  {pages} {042301} (\bibinfo {year} {2017}{\natexlab{b}})},\ \Eprint
  {http://arxiv.org/abs/1703.05808} {arXiv:1703.05808 [nucl-th]} \BibitemShut
  {NoStop}%
\bibitem [{\citenamefont {Alqahtani}\ \emph
  {et~al.}(2017{\natexlab{c}})\citenamefont {Alqahtani}, \citenamefont
  {Nopoush}, \citenamefont {Ryblewski},\ and\ \citenamefont
  {Strickland}}]{Alqahtani:2017tnq}%
  \BibitemOpen
  \bibfield  {author} {\bibinfo {author} {\bibfnamefont {M.}~\bibnamefont
  {Alqahtani}}, \bibinfo {author} {\bibfnamefont {M.}~\bibnamefont {Nopoush}},
  \bibinfo {author} {\bibfnamefont {R.}~\bibnamefont {Ryblewski}}, \ and\
  \bibinfo {author} {\bibfnamefont {M.}~\bibnamefont {Strickland}},\
  }\href@noop {} {\  (\bibinfo {year} {2017}{\natexlab{c}})},\ \Eprint
  {http://arxiv.org/abs/1705.10191} {arXiv:1705.10191 [nucl-th]} \BibitemShut
  {NoStop}%
\bibitem [{\citenamefont {Heller}\ and\ \citenamefont
  {Spalinski}(2015)}]{Heller:2015dha}%
  \BibitemOpen
  \bibfield  {author} {\bibinfo {author} {\bibfnamefont {M.~P.}\ \bibnamefont
  {Heller}}\ and\ \bibinfo {author} {\bibfnamefont {M.}~\bibnamefont
  {Spalinski}},\ }\href {\doibase 10.1103/PhysRevLett.115.072501} {\bibfield
  {journal} {\bibinfo  {journal} {Phys. Rev. Lett.}\ }\textbf {\bibinfo
  {volume} {115}},\ \bibinfo {pages} {072501} (\bibinfo {year} {2015})},\
  \Eprint {http://arxiv.org/abs/1503.07514} {arXiv:1503.07514 [hep-th]}
  \BibitemShut {NoStop}%
\bibitem [{\citenamefont {Romatschke}(2017)}]{Romatschke:2017vte}%
  \BibitemOpen
  \bibfield  {author} {\bibinfo {author} {\bibfnamefont {P.}~\bibnamefont
  {Romatschke}},\ }\href@noop {} {\  (\bibinfo {year} {2017})},\ \Eprint
  {http://arxiv.org/abs/1704.08699} {arXiv:1704.08699 [hep-th]} \BibitemShut
  {NoStop}%
\bibitem [{\citenamefont {Bemfica}\ \emph {et~al.}(2017)\citenamefont
  {Bemfica}, \citenamefont {Disconzi},\ and\ \citenamefont
  {Noronha}}]{Bemfica:2017wps}%
  \BibitemOpen
  \bibfield  {author} {\bibinfo {author} {\bibfnamefont {F.~S.}\ \bibnamefont
  {Bemfica}}, \bibinfo {author} {\bibfnamefont {M.~M.}\ \bibnamefont
  {Disconzi}}, \ and\ \bibinfo {author} {\bibfnamefont {J.}~\bibnamefont
  {Noronha}},\ }\href@noop {} {\  (\bibinfo {year} {2017})},\ \Eprint
  {http://arxiv.org/abs/1708.06255} {arXiv:1708.06255 [gr-qc]} \BibitemShut
  {NoStop}%
\bibitem [{\citenamefont {Spalinski}(2017)}]{Spalinski:2017mel}%
  \BibitemOpen
  \bibfield  {author} {\bibinfo {author} {\bibfnamefont {M.}~\bibnamefont
  {Spalinski}},\ }\href@noop {} {\  (\bibinfo {year} {2017})},\ \Eprint
  {http://arxiv.org/abs/1708.01921} {arXiv:1708.01921 [hep-th]} \BibitemShut
  {NoStop}%
\bibitem [{\citenamefont {Strickland}(2015)}]{Strickland:2013uga}%
  \BibitemOpen
  \bibfield  {author} {\bibinfo {author} {\bibfnamefont {M.}~\bibnamefont
  {Strickland}},\ }\href {\doibase 10.1007/s12043-015-0972-1} {\bibfield
  {journal} {\bibinfo  {journal} {Pramana}\ }\textbf {\bibinfo {volume} {84}},\
  \bibinfo {pages} {671} (\bibinfo {year} {2015})},\ \Eprint
  {http://arxiv.org/abs/1312.2285} {arXiv:1312.2285 [hep-ph]} \BibitemShut
  {NoStop}%
\bibitem [{\citenamefont {Berges}\ \emph {et~al.}(2004)\citenamefont {Berges},
  \citenamefont {Borsanyi},\ and\ \citenamefont {Wetterich}}]{Berges:2004ce}%
  \BibitemOpen
  \bibfield  {author} {\bibinfo {author} {\bibfnamefont {J.}~\bibnamefont
  {Berges}}, \bibinfo {author} {\bibfnamefont {S.}~\bibnamefont {Borsanyi}}, \
  and\ \bibinfo {author} {\bibfnamefont {C.}~\bibnamefont {Wetterich}},\ }\href
  {\doibase 10.1103/PhysRevLett.93.142002} {\bibfield  {journal} {\bibinfo
  {journal} {Phys. Rev. Lett.}\ }\textbf {\bibinfo {volume} {93}},\ \bibinfo
  {pages} {142002} (\bibinfo {year} {2004})},\ \Eprint
  {http://arxiv.org/abs/hep-ph/0403234} {arXiv:hep-ph/0403234 [hep-ph]}
  \BibitemShut {NoStop}%
\bibitem [{\citenamefont {Bjorken}(1983)}]{Bjorken:1982qr}%
  \BibitemOpen
  \bibfield  {author} {\bibinfo {author} {\bibfnamefont {J.~D.}\ \bibnamefont
  {Bjorken}},\ }\href {\doibase 10.1103/PhysRevD.27.140} {\bibfield  {journal}
  {\bibinfo  {journal} {Phys. Rev.}\ }\textbf {\bibinfo {volume} {D27}},\
  \bibinfo {pages} {140} (\bibinfo {year} {1983})}\BibitemShut {NoStop}%
\bibitem [{\citenamefont {Kovtun}\ and\ \citenamefont
  {Starinets}(2005)}]{Kovtun:2005ev}%
  \BibitemOpen
  \bibfield  {author} {\bibinfo {author} {\bibfnamefont {P.~K.}\ \bibnamefont
  {Kovtun}}\ and\ \bibinfo {author} {\bibfnamefont {A.~O.}\ \bibnamefont
  {Starinets}},\ }\href {\doibase 10.1103/PhysRevD.72.086009} {\bibfield
  {journal} {\bibinfo  {journal} {Phys. Rev.}\ }\textbf {\bibinfo {volume}
  {D72}},\ \bibinfo {pages} {086009} (\bibinfo {year} {2005})},\ \Eprint
  {http://arxiv.org/abs/hep-th/0506184} {arXiv:hep-th/0506184 [hep-th]}
  \BibitemShut {NoStop}%
\bibitem [{\citenamefont {Denicol}\ \emph {et~al.}(2012)\citenamefont
  {Denicol}, \citenamefont {Niemi}, \citenamefont {Moln\'{a}r},\ and\
  \citenamefont {Rischke}}]{Denicol:2012cn}%
  \BibitemOpen
  \bibfield  {author} {\bibinfo {author} {\bibfnamefont {G.~S.}\ \bibnamefont
  {Denicol}}, \bibinfo {author} {\bibfnamefont {H.}~\bibnamefont {Niemi}},
  \bibinfo {author} {\bibfnamefont {E.}~\bibnamefont {Moln\'{a}r}}, \ and\
  \bibinfo {author} {\bibfnamefont {D.~H.}\ \bibnamefont {Rischke}},\ }\href
  {\doibase 10.1103/PhysRevD.85.114047} {\bibfield  {journal} {\bibinfo
  {journal} {Phys. Rev. D}\ }\textbf {\bibinfo {volume} {85}},\ \bibinfo
  {pages} {114047} (\bibinfo {year} {2012})}\BibitemShut {NoStop}%
\bibitem [{\citenamefont {Anderson}\ and\ \citenamefont
  {Witting}(1974)}]{anderson1974relativistic}%
  \BibitemOpen
  \bibfield  {author} {\bibinfo {author} {\bibfnamefont {J.}~\bibnamefont
  {Anderson}}\ and\ \bibinfo {author} {\bibfnamefont {H.}~\bibnamefont
  {Witting}},\ }\href@noop {} {\bibfield  {journal} {\bibinfo  {journal}
  {Physica}\ }\textbf {\bibinfo {volume} {74}},\ \bibinfo {pages} {489}
  (\bibinfo {year} {1974})}\BibitemShut {NoStop}%
\bibitem [{\citenamefont {Muller}(1967)}]{Muller:1967zza}%
  \BibitemOpen
  \bibfield  {author} {\bibinfo {author} {\bibfnamefont {I.}~\bibnamefont
  {Muller}},\ }\href {\doibase 10.1007/BF01326412} {\bibfield  {journal}
  {\bibinfo  {journal} {Z. Phys.}\ }\textbf {\bibinfo {volume} {198}},\
  \bibinfo {pages} {329} (\bibinfo {year} {1967})}\BibitemShut {NoStop}%
\bibitem [{\citenamefont {Israel}(1976)}]{Israel:1976tn}%
  \BibitemOpen
  \bibfield  {author} {\bibinfo {author} {\bibfnamefont {W.}~\bibnamefont
  {Israel}},\ }\href {\doibase 10.1016/0003-4916(76)90064-6} {\bibfield
  {journal} {\bibinfo  {journal} {Annals Phys.}\ }\textbf {\bibinfo {volume}
  {100}},\ \bibinfo {pages} {310} (\bibinfo {year} {1976})}\BibitemShut
  {NoStop}%
\bibitem [{\citenamefont {Israel}\ and\ \citenamefont
  {Stewart}(1979)}]{Israel:1979wp}%
  \BibitemOpen
  \bibfield  {author} {\bibinfo {author} {\bibfnamefont {W.}~\bibnamefont
  {Israel}}\ and\ \bibinfo {author} {\bibfnamefont {J.~M.}\ \bibnamefont
  {Stewart}},\ }\href {\doibase 10.1016/0003-4916(79)90130-1} {\bibfield
  {journal} {\bibinfo  {journal} {Annals Phys.}\ }\textbf {\bibinfo {volume}
  {118}},\ \bibinfo {pages} {341} (\bibinfo {year} {1979})}\BibitemShut
  {NoStop}%
\bibitem [{\citenamefont {Baier}\ \emph {et~al.}(2008)\citenamefont {Baier},
  \citenamefont {Romatschke}, \citenamefont {Son}, \citenamefont {Starinets},\
  and\ \citenamefont {Stephanov}}]{Baier:2007ix}%
  \BibitemOpen
  \bibfield  {author} {\bibinfo {author} {\bibfnamefont {R.}~\bibnamefont
  {Baier}}, \bibinfo {author} {\bibfnamefont {P.}~\bibnamefont {Romatschke}},
  \bibinfo {author} {\bibfnamefont {D.~T.}\ \bibnamefont {Son}}, \bibinfo
  {author} {\bibfnamefont {A.~O.}\ \bibnamefont {Starinets}}, \ and\ \bibinfo
  {author} {\bibfnamefont {M.~A.}\ \bibnamefont {Stephanov}},\ }\href {\doibase
  10.1088/1126-6708/2008/04/100} {\bibfield  {journal} {\bibinfo  {journal}
  {JHEP}\ }\textbf {\bibinfo {volume} {0804}},\ \bibinfo {pages} {100}
  (\bibinfo {year} {2008})},\ \Eprint {http://arxiv.org/abs/0712.2451}
  {arXiv:0712.2451 [hep-th]} \BibitemShut {NoStop}%
\bibitem [{\citenamefont {Landau}\ and\ \citenamefont
  {Lifshitz}(1987)}]{LandauLifshitzFluids}%
  \BibitemOpen
  \bibfield  {author} {\bibinfo {author} {\bibfnamefont {L.~D.}\ \bibnamefont
  {Landau}}\ and\ \bibinfo {author} {\bibfnamefont {E.~M.}\ \bibnamefont
  {Lifshitz}},\ }\href@noop {} {\emph {\bibinfo {title} {Fluid Mechanics -
  Volume 6 (Corse of Theoretical Physics)}}}\ (\bibinfo  {publisher} {Pergamon
  Press},\ \bibinfo {year} {1987})\BibitemShut {NoStop}%
\bibitem [{\citenamefont {Denicol}\ \emph {et~al.}(2010)\citenamefont
  {Denicol}, \citenamefont {Koide},\ and\ \citenamefont
  {Rischke}}]{Denicol:2010xn}%
  \BibitemOpen
  \bibfield  {author} {\bibinfo {author} {\bibfnamefont {G.}~\bibnamefont
  {Denicol}}, \bibinfo {author} {\bibfnamefont {T.}~\bibnamefont {Koide}}, \
  and\ \bibinfo {author} {\bibfnamefont {D.}~\bibnamefont {Rischke}},\ }\href
  {\doibase 10.1103/PhysRevLett.105.162501} {\bibfield  {journal} {\bibinfo
  {journal} {Phys.Rev.Lett.}\ }\textbf {\bibinfo {volume} {105}},\ \bibinfo
  {pages} {162501} (\bibinfo {year} {2010})},\ \Eprint
  {http://arxiv.org/abs/1004.5013} {arXiv:1004.5013 [nucl-th]} \BibitemShut
  {NoStop}%
\bibitem [{\citenamefont {Denicol}\ \emph {et~al.}(2011)\citenamefont
  {Denicol}, \citenamefont {Noronha}, \citenamefont {Niemi},\ and\
  \citenamefont {Rischke}}]{Denicol:2011fa}%
  \BibitemOpen
  \bibfield  {author} {\bibinfo {author} {\bibfnamefont {G.~S.}\ \bibnamefont
  {Denicol}}, \bibinfo {author} {\bibfnamefont {J.}~\bibnamefont {Noronha}},
  \bibinfo {author} {\bibfnamefont {H.}~\bibnamefont {Niemi}}, \ and\ \bibinfo
  {author} {\bibfnamefont {D.~H.}\ \bibnamefont {Rischke}},\ }\href {\doibase
  10.1103/PhysRevD.83.074019} {\bibfield  {journal} {\bibinfo  {journal} {Phys.
  Rev.}\ }\textbf {\bibinfo {volume} {D83}},\ \bibinfo {pages} {074019}
  (\bibinfo {year} {2011})},\ \Eprint {http://arxiv.org/abs/1102.4780}
  {arXiv:1102.4780 [hep-th]} \BibitemShut {NoStop}%
\bibitem [{\citenamefont {Denicol}\ \emph
  {et~al.}(2014{\natexlab{a}})\citenamefont {Denicol}, \citenamefont {Heinz},
  \citenamefont {Martinez}, \citenamefont {Noronha},\ and\ \citenamefont
  {Strickland}}]{Denicol:2014xca}%
  \BibitemOpen
  \bibfield  {author} {\bibinfo {author} {\bibfnamefont {G.~S.}\ \bibnamefont
  {Denicol}}, \bibinfo {author} {\bibfnamefont {U.~W.}\ \bibnamefont {Heinz}},
  \bibinfo {author} {\bibfnamefont {M.}~\bibnamefont {Martinez}}, \bibinfo
  {author} {\bibfnamefont {J.}~\bibnamefont {Noronha}}, \ and\ \bibinfo
  {author} {\bibfnamefont {M.}~\bibnamefont {Strickland}},\ }\href {\doibase
  10.1103/PhysRevLett.113.202301} {\bibfield  {journal} {\bibinfo  {journal}
  {Phys. Rev. Lett.}\ }\textbf {\bibinfo {volume} {113}},\ \bibinfo {pages}
  {202301} (\bibinfo {year} {2014}{\natexlab{a}})},\ \Eprint
  {http://arxiv.org/abs/1408.5646} {arXiv:1408.5646 [hep-ph]} \BibitemShut
  {NoStop}%
\bibitem [{\citenamefont {Denicol}\ \emph
  {et~al.}(2014{\natexlab{b}})\citenamefont {Denicol}, \citenamefont {Heinz},
  \citenamefont {Martinez}, \citenamefont {Noronha},\ and\ \citenamefont
  {Strickland}}]{Denicol:2014tha}%
  \BibitemOpen
  \bibfield  {author} {\bibinfo {author} {\bibfnamefont {G.~S.}\ \bibnamefont
  {Denicol}}, \bibinfo {author} {\bibfnamefont {U.~W.}\ \bibnamefont {Heinz}},
  \bibinfo {author} {\bibfnamefont {M.}~\bibnamefont {Martinez}}, \bibinfo
  {author} {\bibfnamefont {J.}~\bibnamefont {Noronha}}, \ and\ \bibinfo
  {author} {\bibfnamefont {M.}~\bibnamefont {Strickland}},\ }\href {\doibase
  10.1103/PhysRevD.90.125026} {\bibfield  {journal} {\bibinfo  {journal} {Phys.
  Rev.}\ }\textbf {\bibinfo {volume} {D90}},\ \bibinfo {pages} {125026}
  (\bibinfo {year} {2014}{\natexlab{b}})},\ \Eprint
  {http://arxiv.org/abs/1408.7048} {arXiv:1408.7048 [hep-ph]} \BibitemShut
  {NoStop}%
\bibitem [{\citenamefont {Denicol}(2014)}]{Denicol:2014loa}%
  \BibitemOpen
  \bibfield  {author} {\bibinfo {author} {\bibfnamefont {G.~S.}\ \bibnamefont
  {Denicol}},\ }\href {\doibase 10.1088/0954-3899/41/12/124004} {\bibfield
  {journal} {\bibinfo  {journal} {J. Phys.}\ }\textbf {\bibinfo {volume}
  {G41}},\ \bibinfo {pages} {124004} (\bibinfo {year} {2014})}\BibitemShut
  {NoStop}%
\bibitem [{\citenamefont {Jaiswal}(2013{\natexlab{a}})}]{Jaiswal:2013vta}%
  \BibitemOpen
  \bibfield  {author} {\bibinfo {author} {\bibfnamefont {A.}~\bibnamefont
  {Jaiswal}},\ }\href {\doibase 10.1103/PhysRevC.88.021903} {\bibfield
  {journal} {\bibinfo  {journal} {Phys. Rev.}\ }\textbf {\bibinfo {volume}
  {C88}},\ \bibinfo {pages} {021903} (\bibinfo {year} {2013}{\natexlab{a}})},\
  \Eprint {http://arxiv.org/abs/1305.3480} {arXiv:1305.3480 [nucl-th]}
  \BibitemShut {NoStop}%
\bibitem [{\citenamefont {Jaiswal}(2013{\natexlab{b}})}]{Jaiswal:2013npa}%
  \BibitemOpen
  \bibfield  {author} {\bibinfo {author} {\bibfnamefont {A.}~\bibnamefont
  {Jaiswal}},\ }\href {\doibase 10.1103/PhysRevC.87.051901} {\bibfield
  {journal} {\bibinfo  {journal} {Phys. Rev.}\ }\textbf {\bibinfo {volume}
  {C87}},\ \bibinfo {pages} {051901} (\bibinfo {year} {2013}{\natexlab{b}})},\
  \Eprint {http://arxiv.org/abs/1302.6311} {arXiv:1302.6311 [nucl-th]}
  \BibitemShut {NoStop}%
\bibitem [{\citenamefont {Muronga}(2004)}]{Muronga:2003ta}%
  \BibitemOpen
  \bibfield  {author} {\bibinfo {author} {\bibfnamefont {A.}~\bibnamefont
  {Muronga}},\ }\href {\doibase 10.1103/PhysRevC.69.034903} {\bibfield
  {journal} {\bibinfo  {journal} {Phys. Rev.}\ }\textbf {\bibinfo {volume}
  {C69}},\ \bibinfo {pages} {034903} (\bibinfo {year} {2004})},\ \Eprint
  {http://arxiv.org/abs/nucl-th/0309055} {arXiv:nucl-th/0309055} \BibitemShut
  {NoStop}%
\bibitem [{\citenamefont {Romatschke}\ and\ \citenamefont
  {Strickland}(2003)}]{Romatschke:2003ms}%
  \BibitemOpen
  \bibfield  {author} {\bibinfo {author} {\bibfnamefont {P.}~\bibnamefont
  {Romatschke}}\ and\ \bibinfo {author} {\bibfnamefont {M.}~\bibnamefont
  {Strickland}},\ }\href {\doibase 10.1103/PhysRevD.68.036004} {\bibfield
  {journal} {\bibinfo  {journal} {Phys. Rev.}\ }\textbf {\bibinfo {volume}
  {D68}},\ \bibinfo {pages} {036004} (\bibinfo {year} {2003})},\ \Eprint
  {http://arxiv.org/abs/hep-ph/0304092} {arXiv:hep-ph/0304092 [hep-ph]}
  \BibitemShut {NoStop}%
\bibitem [{\citenamefont {Romatschke}\ and\ \citenamefont
  {Strickland}(2004)}]{Romatschke:2004jh}%
  \BibitemOpen
  \bibfield  {author} {\bibinfo {author} {\bibfnamefont {P.}~\bibnamefont
  {Romatschke}}\ and\ \bibinfo {author} {\bibfnamefont {M.}~\bibnamefont
  {Strickland}},\ }\href {\doibase 10.1103/PhysRevD.70.116006} {\bibfield
  {journal} {\bibinfo  {journal} {Phys. Rev.}\ }\textbf {\bibinfo {volume}
  {D70}},\ \bibinfo {pages} {116006} (\bibinfo {year} {2004})},\ \Eprint
  {http://arxiv.org/abs/hep-ph/0406188} {arXiv:hep-ph/0406188 [hep-ph]}
  \BibitemShut {NoStop}%
\bibitem [{\citenamefont {Rebhan}\ \emph {et~al.}(2008)\citenamefont {Rebhan},
  \citenamefont {Strickland},\ and\ \citenamefont {Attems}}]{Rebhan:2008uj}%
  \BibitemOpen
  \bibfield  {author} {\bibinfo {author} {\bibfnamefont {A.}~\bibnamefont
  {Rebhan}}, \bibinfo {author} {\bibfnamefont {M.}~\bibnamefont {Strickland}},
  \ and\ \bibinfo {author} {\bibfnamefont {M.}~\bibnamefont {Attems}},\ }\href
  {\doibase 10.1103/PhysRevD.78.045023} {\bibfield  {journal} {\bibinfo
  {journal} {Phys. Rev.}\ }\textbf {\bibinfo {volume} {D78}},\ \bibinfo {pages}
  {045023} (\bibinfo {year} {2008})},\ \Eprint {http://arxiv.org/abs/0802.1714}
  {arXiv:0802.1714 [hep-ph]} \BibitemShut {NoStop}%
\bibitem [{\citenamefont {Romatschke}(2012)}]{Romatschke:2011qp}%
  \BibitemOpen
  \bibfield  {author} {\bibinfo {author} {\bibfnamefont {P.}~\bibnamefont
  {Romatschke}},\ }\href {\doibase 10.1103/PhysRevD.85.065012} {\bibfield
  {journal} {\bibinfo  {journal} {Phys.Rev.}\ }\textbf {\bibinfo {volume}
  {D85}},\ \bibinfo {pages} {065012} (\bibinfo {year} {2012})},\ \Eprint
  {http://arxiv.org/abs/1108.5561} {arXiv:1108.5561 [gr-qc]} \BibitemShut
  {NoStop}%
\bibitem [{\citenamefont {Ling}(2013)}]{Ling:PC}%
  \BibitemOpen
  \bibfield  {author} {\bibinfo {author} {\bibfnamefont {B.}~\bibnamefont
  {Ling}},\ }\href@noop {} {}\bibinfo {howpublished} {Private Communication}
  (\bibinfo {year} {2013})\BibitemShut {NoStop}%
\bibitem [{\citenamefont {Janik}\ and\ \citenamefont
  {Peschanski}(2006)}]{Janik:2005zt}%
  \BibitemOpen
  \bibfield  {author} {\bibinfo {author} {\bibfnamefont {R.~A.}\ \bibnamefont
  {Janik}}\ and\ \bibinfo {author} {\bibfnamefont {R.~B.}\ \bibnamefont
  {Peschanski}},\ }\href {\doibase 10.1103/PhysRevD.73.045013} {\bibfield
  {journal} {\bibinfo  {journal} {Phys. Rev.}\ }\textbf {\bibinfo {volume}
  {D73}},\ \bibinfo {pages} {045013} (\bibinfo {year} {2006})},\ \Eprint
  {http://arxiv.org/abs/hep-th/0512162} {arXiv:hep-th/0512162 [hep-th]}
  \BibitemShut {NoStop}%
\bibitem [{\citenamefont {Hiscock}\ and\ \citenamefont
  {Lindblom}(1983)}]{Hiscock:1983zz}%
  \BibitemOpen
  \bibfield  {author} {\bibinfo {author} {\bibfnamefont {W.~A.}\ \bibnamefont
  {Hiscock}}\ and\ \bibinfo {author} {\bibfnamefont {L.}~\bibnamefont
  {Lindblom}},\ }\href {\doibase 10.1016/0003-4916(83)90288-9} {\bibfield
  {journal} {\bibinfo  {journal} {Annals Phys.}\ }\textbf {\bibinfo {volume}
  {151}},\ \bibinfo {pages} {466} (\bibinfo {year} {1983})}\BibitemShut
  {NoStop}%
\bibitem [{\citenamefont {Pu}\ \emph {et~al.}(2010)\citenamefont {Pu},
  \citenamefont {Koide},\ and\ \citenamefont {Rischke}}]{Pu:2009fj}%
  \BibitemOpen
  \bibfield  {author} {\bibinfo {author} {\bibfnamefont {S.}~\bibnamefont
  {Pu}}, \bibinfo {author} {\bibfnamefont {T.}~\bibnamefont {Koide}}, \ and\
  \bibinfo {author} {\bibfnamefont {D.~H.}\ \bibnamefont {Rischke}},\ }\href
  {\doibase 10.1103/PhysRevD.81.114039} {\bibfield  {journal} {\bibinfo
  {journal} {Phys. Rev.}\ }\textbf {\bibinfo {volume} {D81}},\ \bibinfo {pages}
  {114039} (\bibinfo {year} {2010})},\ \Eprint {http://arxiv.org/abs/0907.3906}
  {arXiv:0907.3906 [hep-ph]} \BibitemShut {NoStop}%
\bibitem [{\citenamefont {Liddle}\ \emph {et~al.}(1994)\citenamefont {Liddle},
  \citenamefont {Parsons},\ and\ \citenamefont {Barrow}}]{Liddle:1994dx}%
  \BibitemOpen
  \bibfield  {author} {\bibinfo {author} {\bibfnamefont {A.~R.}\ \bibnamefont
  {Liddle}}, \bibinfo {author} {\bibfnamefont {P.}~\bibnamefont {Parsons}}, \
  and\ \bibinfo {author} {\bibfnamefont {J.~D.}\ \bibnamefont {Barrow}},\
  }\href {\doibase 10.1103/PhysRevD.50.7222} {\bibfield  {journal} {\bibinfo
  {journal} {Phys. Rev.}\ }\textbf {\bibinfo {volume} {D50}},\ \bibinfo {pages}
  {7222} (\bibinfo {year} {1994})},\ \Eprint
  {http://arxiv.org/abs/astro-ph/9408015} {arXiv:astro-ph/9408015 [astro-ph]}
  \BibitemShut {NoStop}%
\bibitem [{\citenamefont {Florkowski}\ \emph {et~al.}(2016)\citenamefont
  {Florkowski}, \citenamefont {Ryblewski},\ and\ \citenamefont
  {Spalinski}}]{Florkowski:2016zsi}%
  \BibitemOpen
  \bibfield  {author} {\bibinfo {author} {\bibfnamefont {W.}~\bibnamefont
  {Florkowski}}, \bibinfo {author} {\bibfnamefont {R.}~\bibnamefont
  {Ryblewski}}, \ and\ \bibinfo {author} {\bibfnamefont {M.}~\bibnamefont
  {Spalinski}},\ }\href {\doibase 10.1103/PhysRevD.94.114025} {\bibfield
  {journal} {\bibinfo  {journal} {Phys. Rev.}\ }\textbf {\bibinfo {volume}
  {D94}},\ \bibinfo {pages} {114025} (\bibinfo {year} {2016})},\ \Eprint
  {http://arxiv.org/abs/1608.07558} {arXiv:1608.07558 [nucl-th]} \BibitemShut
  {NoStop}%
\bibitem [{\citenamefont {Florkowski}\ \emph
  {et~al.}(2013{\natexlab{a}})\citenamefont {Florkowski}, \citenamefont
  {Ryblewski},\ and\ \citenamefont {Strickland}}]{Florkowski:2013lza}%
  \BibitemOpen
  \bibfield  {author} {\bibinfo {author} {\bibfnamefont {W.}~\bibnamefont
  {Florkowski}}, \bibinfo {author} {\bibfnamefont {R.}~\bibnamefont
  {Ryblewski}}, \ and\ \bibinfo {author} {\bibfnamefont {M.}~\bibnamefont
  {Strickland}},\ }\href {\doibase 10.1016/j.nuclphysa.2013.08.004} {\bibfield
  {journal} {\bibinfo  {journal} {Nucl. Phys.}\ }\textbf {\bibinfo {volume}
  {A916}},\ \bibinfo {pages} {249} (\bibinfo {year} {2013}{\natexlab{a}})},\
  \Eprint {http://arxiv.org/abs/1304.0665} {arXiv:1304.0665 [nucl-th]}
  \BibitemShut {NoStop}%
\bibitem [{\citenamefont {Florkowski}\ \emph
  {et~al.}(2013{\natexlab{b}})\citenamefont {Florkowski}, \citenamefont
  {Ryblewski},\ and\ \citenamefont {Strickland}}]{Florkowski:2013lya}%
  \BibitemOpen
  \bibfield  {author} {\bibinfo {author} {\bibfnamefont {W.}~\bibnamefont
  {Florkowski}}, \bibinfo {author} {\bibfnamefont {R.}~\bibnamefont
  {Ryblewski}}, \ and\ \bibinfo {author} {\bibfnamefont {M.}~\bibnamefont
  {Strickland}},\ }\href {\doibase 10.1103/PhysRevC.88.024903} {\bibfield
  {journal} {\bibinfo  {journal} {Phys. Rev.}\ }\textbf {\bibinfo {volume}
  {C88}},\ \bibinfo {pages} {024903} (\bibinfo {year} {2013}{\natexlab{b}})},\
  \Eprint {http://arxiv.org/abs/1305.7234} {arXiv:1305.7234 [nucl-th]}
  \BibitemShut {NoStop}%
\bibitem [{\citenamefont {Strickland}(2017)}]{MikeCodeDB}%
  \BibitemOpen
  \bibfield  {author} {\bibinfo {author} {\bibfnamefont {M.}~\bibnamefont
  {Strickland}},\ }\href@noop {} {}\bibinfo {howpublished}
  {\url{http://personal.kent.edu/~mstrick6/code/}} (\bibinfo {year}
  {2017})\BibitemShut {NoStop}%
\bibitem [{\citenamefont {Noronha}\ and\ \citenamefont
  {Denicol}(2011)}]{Noronha:2011fi}%
  \BibitemOpen
  \bibfield  {author} {\bibinfo {author} {\bibfnamefont {J.}~\bibnamefont
  {Noronha}}\ and\ \bibinfo {author} {\bibfnamefont {G.~S.}\ \bibnamefont
  {Denicol}},\ }\href@noop {} {\  (\bibinfo {year} {2011})},\ \Eprint
  {http://arxiv.org/abs/1104.2415} {arXiv:1104.2415 [hep-th]} \BibitemShut
  {NoStop}%
\bibitem [{\citenamefont {Heller}\ \emph {et~al.}(2014)\citenamefont {Heller},
  \citenamefont {Janik}, \citenamefont {Spali\`{n}ski},\ and\ \citenamefont
  {Witaszczyk}}]{Heller:2014wfa}%
  \BibitemOpen
  \bibfield  {author} {\bibinfo {author} {\bibfnamefont {M.~P.}\ \bibnamefont
  {Heller}}, \bibinfo {author} {\bibfnamefont {R.~A.}\ \bibnamefont {Janik}},
  \bibinfo {author} {\bibfnamefont {M.}~\bibnamefont {Spali\`{n}ski}}, \ and\
  \bibinfo {author} {\bibfnamefont {P.}~\bibnamefont {Witaszczyk}},\ }\href
  {\doibase 10.1103/PhysRevLett.113.261601} {\bibfield  {journal} {\bibinfo
  {journal} {Phys. Rev. Lett.}\ }\textbf {\bibinfo {volume} {113}},\ \bibinfo
  {pages} {261601} (\bibinfo {year} {2014})},\ \Eprint
  {http://arxiv.org/abs/1409.5087} {arXiv:1409.5087 [hep-th]} \BibitemShut
  {NoStop}%
\bibitem [{\citenamefont {Bazow}\ \emph
  {et~al.}(2015{\natexlab{b}})\citenamefont {Bazow}, \citenamefont {Martinez},\
  and\ \citenamefont {Heinz}}]{Bazow:2015zca}%
  \BibitemOpen
  \bibfield  {author} {\bibinfo {author} {\bibfnamefont {D.}~\bibnamefont
  {Bazow}}, \bibinfo {author} {\bibfnamefont {M.}~\bibnamefont {Martinez}}, \
  and\ \bibinfo {author} {\bibfnamefont {U.~W.}\ \bibnamefont {Heinz}},\
  }\href@noop {} {\  (\bibinfo {year} {2015}{\natexlab{b}})},\ \Eprint
  {http://arxiv.org/abs/1507.06595} {arXiv:1507.06595 [nucl-th]} \BibitemShut
  {NoStop}%
\bibitem [{\citenamefont {Bazow}\ \emph
  {et~al.}(2016{\natexlab{a}})\citenamefont {Bazow}, \citenamefont {Denicol},
  \citenamefont {Heinz}, \citenamefont {Martinez},\ and\ \citenamefont
  {Noronha}}]{Bazow:2015dha}%
  \BibitemOpen
  \bibfield  {author} {\bibinfo {author} {\bibfnamefont {D.}~\bibnamefont
  {Bazow}}, \bibinfo {author} {\bibfnamefont {G.~S.}\ \bibnamefont {Denicol}},
  \bibinfo {author} {\bibfnamefont {U.}~\bibnamefont {Heinz}}, \bibinfo
  {author} {\bibfnamefont {M.}~\bibnamefont {Martinez}}, \ and\ \bibinfo
  {author} {\bibfnamefont {J.}~\bibnamefont {Noronha}},\ }\href {\doibase
  10.1103/PhysRevLett.116.022301} {\bibfield  {journal} {\bibinfo  {journal}
  {Phys. Rev. Lett.}\ }\textbf {\bibinfo {volume} {116}},\ \bibinfo {pages}
  {022301} (\bibinfo {year} {2016}{\natexlab{a}})},\ \Eprint
  {http://arxiv.org/abs/1507.07834} {arXiv:1507.07834 [hep-ph]} \BibitemShut
  {NoStop}%
\bibitem [{\citenamefont {Bazow}\ \emph
  {et~al.}(2016{\natexlab{b}})\citenamefont {Bazow}, \citenamefont {Denicol},
  \citenamefont {Heinz}, \citenamefont {Martinez},\ and\ \citenamefont
  {Noronha}}]{Bazow:2016oky}%
  \BibitemOpen
  \bibfield  {author} {\bibinfo {author} {\bibfnamefont {D.}~\bibnamefont
  {Bazow}}, \bibinfo {author} {\bibfnamefont {G.~S.}\ \bibnamefont {Denicol}},
  \bibinfo {author} {\bibfnamefont {U.}~\bibnamefont {Heinz}}, \bibinfo
  {author} {\bibfnamefont {M.}~\bibnamefont {Martinez}}, \ and\ \bibinfo
  {author} {\bibfnamefont {J.}~\bibnamefont {Noronha}},\ }\href {\doibase
  10.1103/PhysRevD.94.125006} {\bibfield  {journal} {\bibinfo  {journal} {Phys.
  Rev.}\ }\textbf {\bibinfo {volume} {D94}},\ \bibinfo {pages} {125006}
  (\bibinfo {year} {2016}{\natexlab{b}})},\ \Eprint
  {http://arxiv.org/abs/1607.05245} {arXiv:1607.05245 [hep-ph]} \BibitemShut
  {NoStop}%
\end{thebibliography}%

\end{document}